%% file: main.tex
\documentclass{article}

\usepackage{notation}
\usepackage[preprint,nonatbib]{neurips_2025}
\usepackage{etoolbox}
\usepackage{thmtools}
\newcommand{\citep}{\cite}
\newcommand{\citet}{\cite}

\usepackage[utf8]{inputenc} %
\usepackage[T1]{fontenc}    %
\usepackage{hyperref}       %
\usepackage{url}            %
\usepackage{booktabs}       %
\usepackage{amsfonts}       %
\usepackage{nicefrac}       %
\usepackage{microtype}      %
\usepackage[dvipsnames]{xcolor}         %

\usepackage{thmtools}
\usepackage[normalem]{ulem}
\usepackage{wrapfig}
\usepackage{algorithm}
\usepackage{algpseudocode}

\usepackage{enumitem}

\usepackage{chngcntr}
\counterwithout{figure}{section}
\counterwithout{table}{section}

\hypersetup{
    colorlinks=true,
    linkcolor=violet,
    citecolor=magenta
}

\newif\ifenablecomments

\enablecommentstrue

\title{PolyMon: A Unified Framework for Polymer Property Prediction}

\newif\ifgambit
\gambitfalse
\gambittrue

\author{%
Gaopeng Ren\thanks{Equal contribution}\\
Department of Chemistry\\
Imperial College London\\
\And
Yijie Yang\footnotemark[1]\\
Department of Chemistry\\
Imperial College London
\And
Jiajun Zhou\\
Department of Chemistry\\
Imperial College London\\
\AND
Kim E. Jelfs\thanks{Corresponding email: k.jelfs@imperial.ac.uk}\\
Department of Chemistry\\
Imperial College London\\
}

\begin{document}

\maketitle

\setcounter{footnote}{0} 

\begin{abstract}
Accurate prediction of polymer properties is essential for materials design, but remains challenging due to data scarcity, diverse polymer representations, and the lack of systematic evaluation across modelling choices. Here, we present PolyMon, a unified and accessible framework that integrates multiple polymer representations, machine learning methods, and training strategies within a single, accessible platform. PolyMon supports various descriptors and graph construction strategies for polymer representations, and includes a wide range of models, from tabular models to graph neural networks, along with flexible training strategies including multi-fidelity learning, $\Delta$-learning, active learning, and ensemble learning. Using five key polymer properties as benchmarks, we perform systematic evaluations to assess how representations and models affect predictive performance. These case studies further illustrate how different training strategies can be applied within a consistent workflow to leverage limited data and incorporate physical model-derived information. Overall, PolyMon provides a comprehensive and extensible foundation for benchmarking and advancing machine learning-based polymer property prediction. The code is available at \href{https://github.com/fate1997/polymon}{github.com/fate1997/polymon}.
\end{abstract}

\input{section/intro}
\input{section/models}
\input{section/strategy}
\input{section/result}
\input{section/conclusion}

\section*{Author contributions}
G.R. and Y.Y. developed the framework, conducted experiments, and analysed the results. J.Z. contributed to project conceptualisation and expertise in pre-trained polymer language models. K.E.J. supervised the project. G.R. and Y.Y. drafted the manuscript, and all authors contributed to the final version.

\section*{Conflicts of interest}
There are no conflicts of interest to declare.

\section*{Data availability}
All the data, code and models used to generate the results are available in the Github repository at \href{https://github.com/fate1997/polymon}{github.com/fate1997/polymon}.

\section*{Acknowledgements}
G.R. acknowledge Imperial College London for funding through the President's PhD Scholarship. K.E.J. acknowledges the AI for Chemistry: AIchemy hub for funding (EPSRC grant EP/Y028775/1 and EP/Y028759/1). We thank the organisers of the ``NeurIPS – Open Polymer Prediction 2025'' Kaggle competition for providing the polymer datasets and inspiring this work.

\bibliographystyle{unsrt}
\bibliography{references}

\newpage
\appendix

\end{document}


\newpage

\tableofcontents
\clearpage
\newpage

\section{Program details}
\subsection{Arguments, descriptors, and models}
\input{table/arguments}
\input{table/avail_descriptor}
\input{table/avail_model}
\clearpage

\subsection{Example commands for training tabular models}
\textbf{Hyperparameter optimisation}: 
\begin{verbatim}
polymon train \
    --raw-csv ./database/database.csv \
    --sources Kaggle PI1070 PolyMetriX \
    --labels Tg Tc FFV Density Rg \
    --feature-names rdkit2d \
    --model rf \
    --n-fold 5 \
    --n-trials 15 \
    --out-dir ./results \
    --tag debug
\end{verbatim}
\medskip
\textbf{Hyperparameters from a file}: 
\begin{verbatim}
polymon train \
    --raw-csv ./database/database.csv \
    --labels Rg \
    --feature-names ecfp4 \
    --model rf \
    --n-fold 5 \
    --hparams-from ./results/rf/rf-Rg-ecfp4.pkl\
    --out-dir ./results \
    --tag debug
\end{verbatim}

\clearpage

\subsection{Example commands for training neural networks}
\textbf{Hyperparameter optimisation}: 
\begin{verbatim}
polymon train \
    --raw-csv ./database/database.csv \
    --sources Kaggle PI1070 PolyMetriX \
    --labels Tc Tg FFV Density Rg \
    --model gatv2 \
    --n-trials 15 \
    --n-fold 5 \
    --out-dir ./results \
    --tag debug \
    --num-epochs 2500 \
    --early-stopping-patience 250
\end{verbatim}
\medskip
\textbf{Hyperparameters from a file}: 
\begin{verbatim}
polymon train \
    --raw-csv ./database/database.csv \
    --tag debug \
    --labels Rg \
    --model gatv2 \
    --n-fold 5 \
    --hparams-from ./results/gatv2/Rg/paper/hparams_opt/hparams.json \
    --out-dir ./results
\end{verbatim}

\clearpage

\subsection{Example commands for different training strategies}
\subsubsection{Multi-fidelity learning}
\textbf{Finetune the prediction head}: 
\begin{verbatim}
polymon train \
    --tag finetune \
    --labels Density \
    --model gatv2 \
    --n-fold 5 \
    --finetune \
    --finetune-csv-path database/database.csv \
    --sources MAFA-exp \
    --pretrained-model ./results/gatv2/Density/gatv2_Density_pretrained.pt \
    --hparams-from ./results/gatv2/Density/hparams.json
\end{verbatim}
\medskip
\textbf{Label residual}: 
\begin{verbatim}
polymon train \
    --tag label-residual \
    --labels Density \
    --model gatv2 \
    --n-fold 5 \
    --train-residual \
    --sources MAFA-exp \
    --low-fidelity-model ./results/gatv2/Density/gatv2_Density_pretrained.pt \
    --hparams-from ./results/gatv2/Density/hparams.json 
\end{verbatim}
\medskip

\clearpage
\noindent\textbf{Embedding residual}: 
\begin{verbatim}
polymon train \
    --tag emb-residual \
    --labels Density \
    --model gatv2_embed_residual \
    --n-fold 5 \
    --sources MAFA-exp \
    --emb-model ./results/gatv2/Density/gatv2_Density_pretrained.pt \
    --hparams-from ./results/gatv2/Density/hparams.json
\end{verbatim}

\subsubsection{$\Delta$-learning}
\noindent\textbf{Property knowledge transfer}: 
\begin{verbatim}
polymon train \
    --tag from-Tc \
    --labels Density \
    --model gatv2_embed_residual \
    --n-fold 5 \
    --sources PI1070 Kaggle PolyMetriX \
    --emb-model ./results/gatv2/Tc/gatv2_Tc_paper.pt \
    --hparams-from ./results/gatv2/Density/hparams.json
\end{verbatim}
\clearpage

\noindent\textbf{Empirical equations}: 
\begin{verbatim}
polymon train \
    --tag ibm \
    --labels Density \
    --model gatv2 \
    --n-fold 5 \
    --train-residual \
    --sources PI1070 Kaggle PolyMetriX \
    --estimator-name Density-IBM \
    --hparams-from ./results/gatv2/Density/hparams.json
\end{verbatim}

\noindent\textbf{Atomic contributions}: 
\begin{verbatim}
polymon train \
    --tag residual \
    --labels FFV \
    --model gatv2 \
    --n-fold 5 \
    --train-residual \
    --sources PI1070 Kaggle PolyMetriX \
    --hparams-from ./results/gatv2/FFV/hparams.json
\end{verbatim}

\clearpage

\subsubsection{Active learning}
\noindent\textbf{Data acquisition}: 
\begin{verbatim}
polymon rec \
    --pool-csv ./database/raw_csv/JCIM_sup_bigsmiles.csv \
    --trained-model results/mlp/Rg/paper/mlp_Rg_paper-KFold.pt \
    --acquisition uncertainty \
    --save-path debug.csv \
    --sample-size 20
\end{verbatim}

\subsubsection{Ensemble learning}
\begin{verbatim}
polymon train \
    --raw-csv ./database/Rg_base.csv \
    --sources Kaggle PI1070 \
    --labels Rg
    --model gatv2 \
    --hparams-from ./hparams.json \
    --n-estimators 10 \
    --ensemble-type voting \
    --skip-train \
    --run-production \
    --tag voting
\end{verbatim}
\clearpage

\subsection{Example commands for training different graphs}
\textbf{Periodic graph}: 
\begin{verbatim}
polymon train \
    --raw-csv ./database/database.csv \
    --sources Kaggle PI1070 PolyMetriX \
    --labels Tc Tg FFV Density Rg \
    --model gatv2 \
    --n-fold 5 \
    --n-trials 15 \
    --out-dir ./results \
    --tag periodic \
    --additional-features monomer periodic_bond
\end{verbatim}
\medskip
\textbf{Graphs with virtual nodes}: 
\begin{verbatim}
polymon train \
    --raw-csv ./database/database.csv \
    --labels Rg \
    --sources PI1070 Kaggle PolyMetriX \
    --model gatv2vn \
    --n-fold 5 \
    --descriptors rdkit2d \
    --out-dir ./results \
    --tag debug
\end{verbatim}

\clearpage

\subsection{Example Code for Inference}
After training is complete, the output folder may contain three types of model checkpoints: the single model, the K-fold model (filename ending with \texttt{KFold.pt}), and the ensemble model (filename ending with \texttt{ensemble.pt}). Each saved checkpoint includes all the information required for inference, such as model hyperparameters, model parameters, feature names, normalization parameters, and, if applicable, transformation arguments. Consequently, these models can be loaded and used for prediction directly and effortlessly.\\
\\
\textbf{Single model \& K-fold model}: 
\begin{verbatim}
from polymon.model.base import ModelWrapper

model = ModelWrapper.from_file('results/mlp/Rg/paper/mlp_Rg_paper-KFold.pt')
# The output shape will be [N, 1] for single models, and [N, k] for k-fold models.
predictions = model.predict(['*C*', '*CCC*'])
\end{verbatim}
\medskip
\textbf{Ensemble model}: 
\begin{verbatim}
from polymon.model.ensemble import EnsembleModelWrapper

model_path = 'results/gatv2/Rg/bagging/ensemble/production/gatv2_Rg-ensemble.pt'
model = EnsembleModelWrapper.from_file(model_path)
predictions = model.predict(['*C*', '*CC*'])
\end{verbatim}

\clearpage

\section{Hyperparameter tuning}
The model was trained using five-fold cross-validation, with hyperparameter optimisation performed using Optuna\footnote{\url{https://github.com/optuna/optuna}}
 over a total of 15 trials. The maximum number of training epochs was set to 2000, and early stopping based on validation loss was applied to prevent overfitting. Learning rate is set between $10^{-4}$ and $2\times10^{-3}$ at a logrithimc scale.

\subsection{Hyperparameter for tabular models}

\input{table/hparams_tabular}

\clearpage
\subsection{Hyperparameter for GNN models}
\input{table/hparams_gnn}

\clearpage

\section{Additional model details}

\subsection{KAN-based GNNs}

In this work, we implemented KAN-based GNNs using two complementary strategies. The first follows the approach of Li $et$ $al.$~\cite{Li2025}, in which the linear node projection layer is replaced with a Fourier KAN layer~\cite{xu2024fourierkan}. The second strategy is based on FastKAN~\cite{li2024kolmogorovarnoldnetworksradialbasis}, which can be used as a drop-in replacement for the MLP blocks in several GNN architectures, including GIN~\cite{xu2018how}, GPS~\cite{GPS}, and GATv2~\cite{brody2022how}.

\subsection{Multi-fidelity learning}

For multi-fidelity learning, we first optimised the hyperparameters of a GATv2 model on the low-fidelity dataset ($\rho_{\rm md}$) using 15 optimisation trials with 5-fold cross-validation. The optimised hyperparameters were then used to train a production model, in which the dataset was split into training and validation sets with a 95:5 ratio. This procedure resulted in a single pretrained model for density prediction.
Building upon this pretrained model, we explored several multi-fidelity transfer strategies. For fine-tuning, we considered two variants: updating all parameters of the pretrained model (\textbf{finetune (all)}) and updating only the parameters of the prediction head while freezing the remaining layers (\textbf{finetune (freeze)}). 

For the label-residual and embedding-residual strategies, we trained an additional model using the same hyperparameters as the pretrained one. In the label-residual approach, the model was trained to predict the residual between the ground-truth labels and the predictions produced by the pretrained model. In the embedding-residual approach, the graph embeddings generated by the pretrained model were added to the embeddings produced by the current model to guide the learning process.

\subsection{$\Delta$-learning}

\subsubsection{Property knowledge transfer}
Many polymer properties are correlated through their underlying chemical structure and chain architecture. To explore whether such relationships can be leveraged during model training, we introduced a property knowledge transfer strategy that incorporates information learned from one property into the prediction of another. The core idea is to reuse representations learned from a model trained on a related property and integrate this information into the target property model in a controlled manner.

As a case study, we used thermal conductivity (TC) to assist density prediction. Specifically, a GATv2 model was trained on the TC dataset. The learned polymer embeddings from the pre-trained TC model are then added to the polymer representation learned by the density model through a residual connection. If the embedding dimensions of the two models differ, a linear projection is applied to ensure compatibility before combining the representations. The final prediction is made using the combined representation, which is then passed through an multi-layer perceptron (MLP) prediction head.

This property knowledge transfer strategy is fully supported within the framework and can be applied to any pair of related polymer properties. Its impact on predictive performance is assessed empirically, allowing the framework to systematically examine when and how transferring information across properties is beneficial, neutral, or detrimental.

\subsubsection{Empirical equations}
Empirical equations can provide fast and physically motivated estimates of polymer properties. Instead of learning the target property directly, a model can be trained to learn the difference ($\Delta$ value) between the empirical estimate and the ground truth value. However, empirical relationships are not universally available for all polymer properties, and their accuracy can vary across chemical spaces. Here, we incorporate empirical equations for two properties studied in this framework, namely the radius of gyration ($R_{g}$) and density.

For $R_{g}$, we implemented two empirical estimators that derive polymer size from monomer structure and physical assumptions without generating three-dimensional conformations. Both methods estimate the monomer contour length from the molecular graph using bond length information, followed by a polymer scaling factor based on the degree of polymerisation, characteristic ratio, and solvent condition. The difference lies in how they derive monomer contour length from the whole polymer chain molecule. One implementation approximates the longest backbone path using a depth-first search (DFS) over the molecular graph, while the other formulates the molecule as a weighted graph and estimates the backbone length using graph shortest path analysis in NetworkX\footnote{\url{https://github.com/networkx/networkx}}. We used the first strategy to generate the results in the main text. The empirical equation of the $R_{g}$ is written as
\begin{equation}
    \hat{R_g}=\frac{b}{\sqrt{6}}N_k^\nu
\end{equation}
\begin{equation}
    N_K=\frac{L_c}{b}
\end{equation}
\begin{equation}
    L_c=Nl_m
\end{equation}
where $b$ is the Kuhn length and $\nu$ is the solvent dependent scaling exponent. $N_K$ is the number of Kuhn segments, $L_c$ and $l_m$ is the polymer chain and monomer contour length respectively. $N$ is the degree of polymerisation. 

For density, we considered three empirical estimation strategies with different levels of chemical detail. The overall idea is to divide the molecular weight of the monomer by its molecular volume. Two methods differ in how the molar volume is computed. The first method estimates the molar volume from the van der Waals volume of constituent atoms combined with a packing coefficient. The second method follows a group contribution approach developed by IBM\footnote{\url{https://github.com/IBM/polymer_property_prediction/tree/main}}, where the molar volume is inferred from atom connectivity indices and functional group corrections. The third method adopts the Fedors group contribution scheme\footnote{\url{https://github.com/CalebBell/thermo}}, which estimates molar volume by summing contributions from atom types, functional groups, bond types and ring structures. Together, these estimators provide complementary baseline predictions for density.

\subsubsection{Atomic contribution}
Atomic contribution models provide a simple and interpretable way to relate polymer composition to material properties. We implemented an atomic contribution approach that estimates a target property as a linear combination of element-specific contributions weighted by their occurrence in the polymer. In this work, this strategy is applied to thermal conductivity and fractional free volume (FFV).

In this formulation, a polymer is represented by its atomic composition vector, where each entry corresponds to the number of atoms of a given element in the polymer repeating unit. The property value is then estimated as the weighted sum of these atomic counts, with the weights representing element-specific contributions. These contributions are obtained by fitting a linear regression model to available labelled data. During training, the coefficients are learned by minimising the error between predicted and reference property values across the training dataset, with no intercept term to ensure the predictions depend solely on atomic composition.

\subsection{Active learning}

MD simulations were performed to generate labels for selected monomers. We used RadonPy\cite{Hayashi2022}, an automated workflow for all-atom classical MD simulations, to construct amorphous polymer structures from monomer SMILES strings and compute $R_{g}$ values. For each monomer, an amorphous cell containing six polymer chains was generated, with each chain comprising approximately 800 atoms. The reported $R_{g}$ value was computed from MD trajectories using per-chain data output by LAMMPS. Specifically, time-resolved chain-wise $R_{g}$ values were generated along with the simulation trajectory. For each saved timestep, the $R_{g}$ was evaluated for all polymer chains in the system. To ensure equilibration, only the final 2000 timesteps of each trajectory were considered for analysis. The $R_{g}$ was first averaged over time for each individual chain, and the resulting values were then averaged over all chains to obtain a single mean $R_{g}$ for the system. Mathematically, this corresponds to
\begin{equation}
    <R_g>=\frac{1}{N_{chains}}\sum^{N_{chains}}_{c=1}(\frac{1}{N_t}\sum_tR_g(t,c))
\end{equation}
where $R_g(t,c)$ denotes the radius of gyration of chain $c$ at timestep $t$, $N_t$ is the number of timesteps considered, and $N_{chains}$ is the number of polymer chains in the system.

\clearpage

\section{Additional results}
\subsection{Full results of tabular models on different descriptors}
\input{table/tabular_all}
\clearpage

\subsection{Attempts on other models}

In addition to the models discussed in the main text, we explored several alternative modelling strategies in this work. We report these results to provide additional insights for the polymer community on network design choices for polymer property prediction.

One approach is to explicitly incorporate polymer chain representations into the model. To construct a chain-level representation, we replicated the monomer embedding $n$ times, where $n$ denotes the polymer chain length. A reversal-invariant positional encoding is then applied to differentiate monomer positions along the chain. A class token is prepended to the sequence, and a three-layer Transformer encoder is used to update the monomer representations. We refer to this model as \texttt{gatv2\_chain\_readout}.

Another approach is motivated by the observation that atom ordering within a monomer is not strictly permutation-invariant due to the presence of attachment points. To account for this, we defined atom location features based on the shortest-path distance from each atom to the nearest attachment site. These location features are further encoded using positional encodings and added to the initial atom embeddings. This variant is denoted as \texttt{gatv2\_pe}.

We also evaluated the Directed Message Passing Neural Network (DMPNN) \cite{Yang2019} as a representative message-passing baseline.

The performance of these three models is summarised in Table~\ref{tab:si_gnn}. The results indicate that \texttt{gatv2\_chain\_readout} yields improved performance for predicting $R_{g}$, which is consistent with the fact that $R_{g}$ is inherently a chain-level property. The \texttt{gatv2\_pe} model achieves performance comparable to, but not exceeding, that of the vanilla GATv2. DMPNN attains strong performance on density and TC; however, due to its substantially longer training time, experiments for FFV and $T_{g}$ could not be completed within the allocated 72-hour time limit.

\begin{table}[htbp]
\centering
\caption{Performance of additional GNN models on polymer property prediction (* indicates that training was not completed within 72 hours).}
\label{tab:si_gnn}
\resizebox{\textwidth}{!}{
\renewcommand{\arraystretch}{1.2}
\begin{tabular}{lccccc}
\hline
Model & $\rho_{\rm md}$ & FFV & $R_{g}$ & TC & $T_{g}$ \\
\hline
GATv2 &
$0.0136_{\pm 0.0005}$ &
$0.0046_{\pm 0.0001}$ &
$1.7065_{\pm 0.0711}$ &
$0.0172_{\pm 0.0007}$ &
$22.4999_{\pm 0.7899}$ \\

\texttt{gatv2\_chain\_readout} &
$0.0166_{\pm 0.0010}$ &
* &
$1.6924_{\pm 0.0852}$ &
$0.0186_{\pm 0.0018}$ &
* \\

\texttt{gatv2\_pe} &
$0.0143_{\pm 0.0007}$ &
$0.0048_{\pm 0.0002}$ &
$1.8040_{\pm 0.0953}$ &
$0.0181_{\pm 0.0026}$ &
$25.9572_{\pm 0.9894}$ \\

DMPNN &
$0.0114_{\pm 0.0006}$ &
* &
$1.7144_{\pm 0.0926}$ &
$0.0157_{\pm 0.0019}$ &
* \\
\hline
\end{tabular}}
\end{table}

\clearpage

\subsection{Performance of empirical equations and atomic contributions}
\begin{figure*}[h]
    \centering
	\includegraphics[width=1.0\linewidth]{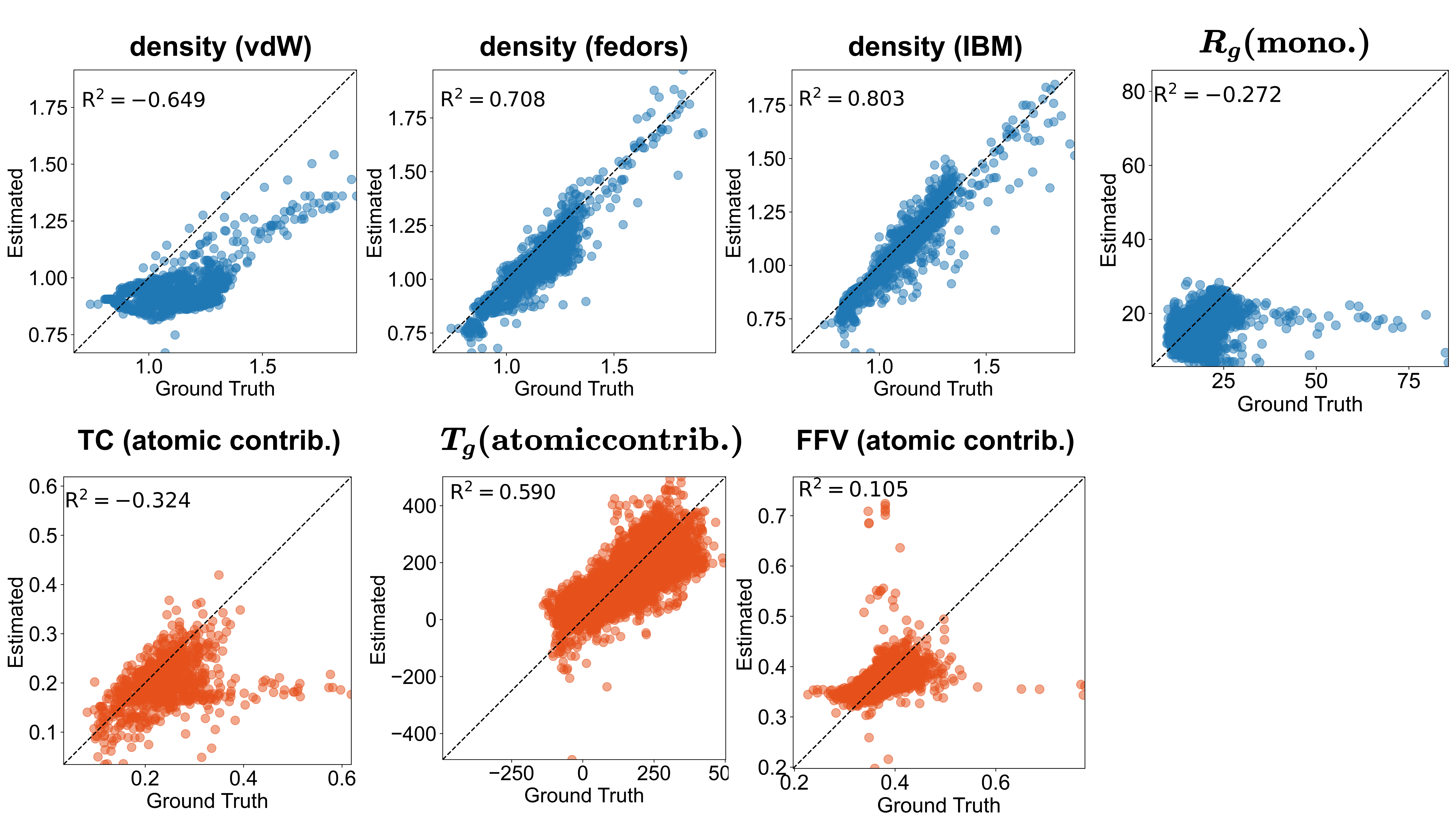}
	\caption{Performance of different empirical equations and atomic contribution methods. }
	\label{fig:estimator_performance}
\end{figure*}

We estimated the target properties using empirical equations for the corresponding datasets (all sets) and evaluated their performance using parity plots against the ground-truth values (Figure~\ref{fig:estimator_performance}). For density estimation, the Fedors and IBM group contribution methods perform well, achieving $R^2$ values greater than 0.7. In contrast, among the atomic contribution methods, good performance is observed for $T_g$, whereas the predictions for TC and FFV are poor.

\clearpage

\bibliography{references}

%% file: section/intro.tex
\section{Introduction}
Polymers underpin a wide range of industrial applications, including therapeutics \cite{Haag2006}, sensors \cite{Adhikari2004}, semiconductors \cite{Ding2023}, agriculture \cite{Cherwoo2024}, and energy technologies \cite{Bose2011}. Accurate prediction of polymer properties can enable large-scale virtual screening and inverse design. Machine learning (ML) has become an important tool for polymer property prediction \cite{Tao2021, Patra2022, Yan2023}. Similar to small organic molecules, polymers can be represented using descriptors, molecular graphs, or sequence-based formats. Descriptor-based methods provide efficient baseline models for polymer property prediction \cite{1650254, D3PY00395G, Bradford2023}. Graph neural networks (GNNs) have attracted increasing attention for learning representations from molecular graphs and have shown promising performance across a range of polymer properties \cite{Queen2023, D2SC02839E, Park2022, Antoniuk2022, Gurnani2023}. Sequence-based representations, such as SMILES strings, further enable large language models to learn transferable polymer representations through unsupervised pretraining on large-scale unlabeled data \cite{D4DD00236A, Han2024, D3SC05079C, Xu2023, Kuenneth2023, Liu2025}.

Data scarcity remains one of the central challenges in training ML models for polymer property prediction. To mitigate this limitation, a range of training strategies has been proposed, including multi-fidelity learning, $\Delta$-learning, active learning, and ensemble learning. Multi-fidelity learning seeks to exploit both high-fidelity experimental data and lower-fidelity computational data \cite{PATRA2020109286, Fare2022, Buterez2024, Ko2025}, which is particularly relevant for polymer systems where experimental measurements are scarce and molecular dynamics (MD) simulations offer a more accessible source of labelled data \cite{Hayashi2022}. Low-fidelity information can be incorporated through techniques such as transfer learning and $\Delta$-learning. $\Delta$-learning, which aims to learn the residual corrections for labels or embeddings, has gained increasing attention, especially for energy-related properties \cite{Ramakrishnan2015, Shen2016, Huang_2025}. From a data-centric perspective, active learning provides an efficient strategy to selectively acquire the most informative data points for improving model performance \cite{D4DD00267A, Xu2024, Jablonka2021, Jose2024}. Ensemble learning improves robustness by aggregating predictions from multiple models. Despite these advances, systematic investigations into how polymer representations, model architectures, and training strategies jointly influence predictive performance remain limited. Moreover, several recently developed ML techniques, such as TabPFN \cite{Hollmann2025}, Kolmogorov-Arnold networks (KANs) \cite{liu2025kan}, and modern $\Delta$-learning formulations, are still relatively underexplored in polymer property prediction.

Here, we present PolyMon, a unified and accessible framework that integrates diverse polymer representations, multiple ML models, and flexible training strategies. PolyMon supports polymer featurisation using descriptors from heterogeneous sources as well as molecular graphs constructed with different strategies. The framework incorporates a broad range of ML models, including tabular learners and graph neural networks (GNNs), and further includes recently proposed architectures such as KANs and KAN-based GNNs. In addition, PolyMon provides easy-to-use implementations of advanced training strategies, including multi-fidelity learning, $\Delta$-learning, active learning, and ensemble learning. Leveraging this framework, we conduct extensive experiments on polymer property prediction using five key polymer properties as benchmarks.


%% file: section/models.tex
\section{Polymer Representations and Models}

\begin{figure}[h]
    \centering
	\includegraphics[width=0.8\linewidth]{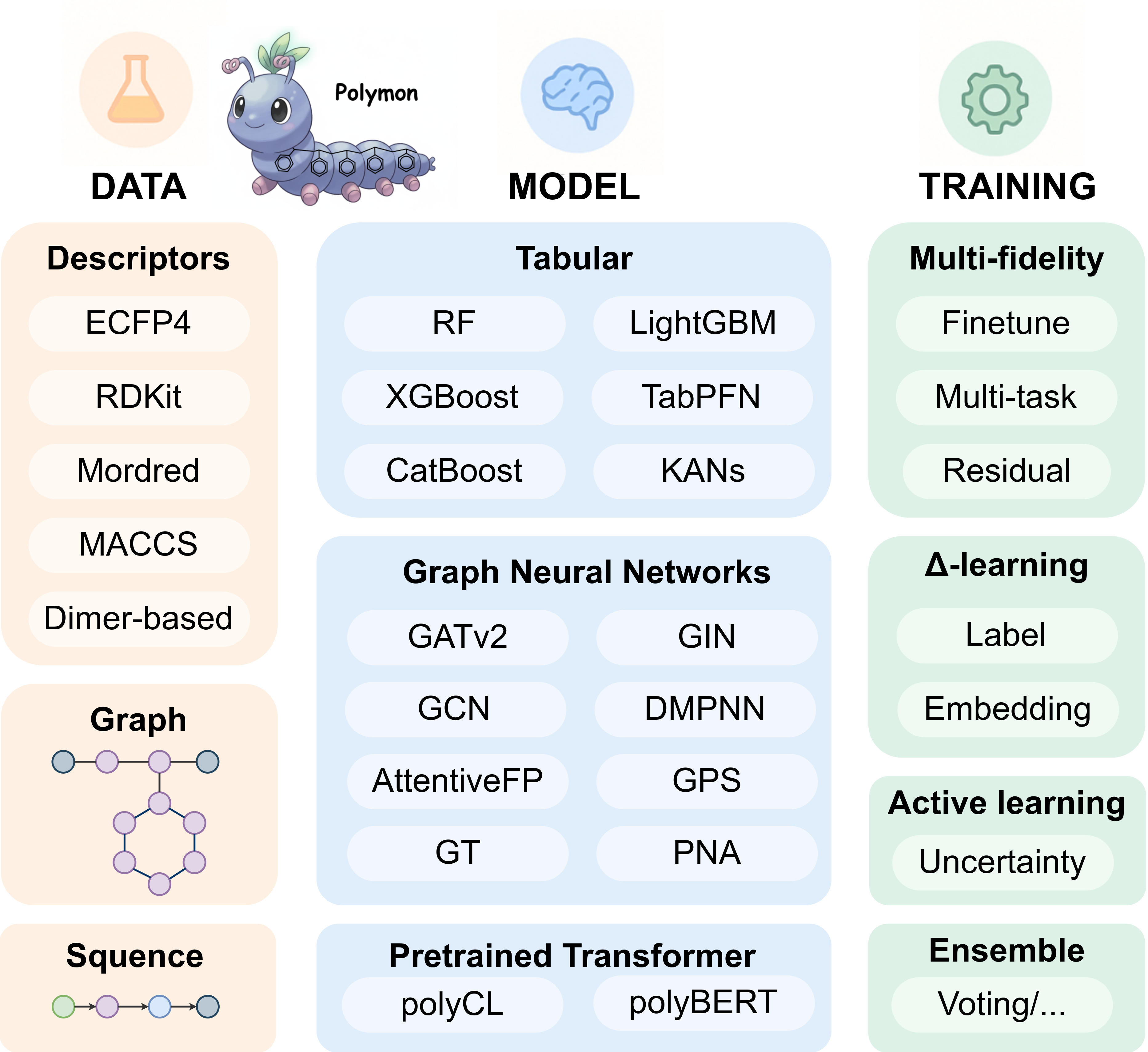}
	\caption{An overview of the PolyMon framework. The framework supports multiple polymer representations, modelling approaches, and training strategies for polymer property prediction. The PolyMon logo was designed with assistance from Gemini.}
	\label{fig:framework}
\end{figure}

The overall PolyMon framework is illustrated in Figure~\ref{fig:framework}. PolyMon supports multiple data representations, ML models, and training strategies, and all experiments in this work are conducted using it. Training and evaluation across different data types, models, and strategies can be performed via a single command-line script. The available command arguments and example codes used to generate the results reported here are provided in Section~S1 of the Supplementary Information (SI). To ensure fair comparisons between models, we perform hyperparameter optimisation for each model, with the hyperparameter search spaces listed in Section~S2 of the SI.

\subsection{Descriptors and Tabular Models}
Polymer descriptors are numerical representations that can be used as input for tabular ML models. Directly calculating descriptors for polymers is often challenging, as the exact degree of polymerisation is typically unknown. Consequently, we focus primarily on descriptors derived from monomers and dimers and use them to approximate polymer representations.  

Extended-connectivity fingerprints (ECFPs) \cite{rogers2010extended} are widely used structural descriptors for molecules. These circular topological fingerprints are represented as binary vectors. Here, we use ECFP with a radius of 2 and a length of 2048, referred to as \textbf{ECFP4}. The Molecular ACCess System (MACCS) keys \cite{Durant2002} are 166-bit binary vectors representing the presence of predefined substructures. Both ECFP and MACCS descriptors are calculated using RDKit \cite{rdkit}. RDKit also provides another predefined descriptor set (\textbf{RDKit}) that includes common physicochemical properties (e.g., molecular weight) and structural features (e.g., number of rotatable bonds). We removed descriptors that produced "NaN" or infinite values, yielding a total of 196 usable RDKit descriptors.  \textbf{Mordred} \cite{Moriwaki2018} is a specialised software for descriptor calculation that offers a wide range of structural and property descriptors. After filtering unstable descriptors, we obtained 704 usable Mordred descriptors. As polymers consist of repeating units, descriptors derived from a single monomer may miss information about interactions between units. To capture such information, we also calculated descriptors for dimers, denoted as \textbf{ECFP (dimer)}, \textbf{RDKit (dimer)}, and \textbf{Mordred (dimer)}.  Large language models provide an alternative approach for generating polymer representations. Here, we evaluated descriptors from \textbf{PolyBERT} \cite{Kuenneth2023} and our prior work \textbf{PolyCL} \cite{D4DD00236A}, both of which were trained on millions of polymers using unsupervised learning, with descriptor vectors of length 600.

With these descriptors, tabular models can be used to fit the data and make predictions. Here, we evaluated several tree-based methods, including random forest (\textbf{RF}) \cite{Breiman2001}, \textbf{XGBoost} \cite{Chen2016}, \textbf{CatBoost} \cite{catboost}, and \textbf{LightGBM} \cite{lightgbm}. We also tested a tabular foundation model, \textbf{TabPFN} \cite{Hollmann2025}, which is pretrained on synthesised tabular datasets. Multi-layer perceptrons (\textbf{MLP}) \cite{HORNIK1989359}, a fundamental module in deep learning, can also be applied to tabular inputs.  Recently, Kolmogorov-Arnold Networks (KANs) \cite{liu2025kan} have emerged as a promising alternative to MLPs. However, the original KANs have a slow training process, making them impractical for datasets with hundreds of descriptors. To address this, we selected several KAN variants with substantially faster training, including \textbf{FastKAN} \cite{li2024kolmogorovarnoldnetworksradialbasis}, \textbf{FourierKAN} \cite{xu2024fourierkan}, and \textbf{EfficientKAN} \cite{Blealtan2024}.

\subsection{Graph and GNNs}

\begin{figure*}[h]
    \centering
	\includegraphics[width=0.45\linewidth]{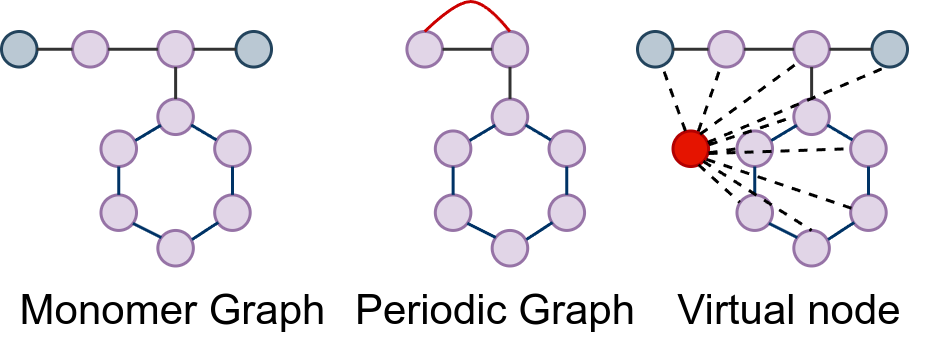}
	\caption{Polymer graphs. Purple nodes denote atoms, blue nodes represent the neighbours of attachment points, and the red node indicates a virtual node. Black edges correspond to chemical bonds, red edges denote additional connections between attachment points, and dashed edges represent virtual connections between atoms and the virtual node.}
	\label{fig:graph}
\end{figure*}

Molecular graphs can capture intrinsic information about molecular structures. Polymers can be represented using several types of graphs (Figure~\ref{fig:graph}). We considered three graph representations: (i) \textbf{monomer graphs}, in which neighbours of attachment points are treated as special atoms; (ii) \textbf{periodic graphs}, which introduce a special edge connecting attachment points to capture repeating structures; and (iii) \textbf{graphs with virtual nodes}, where a virtual node is added and connected to all other nodes, with the final representation of the graph taken from the virtual node embedding. Unless otherwise noted, we primarily use monomer graphs in this study.

GNNs can be applied to these graphs to iteratively update node (and optionally edge) embeddings through multiple message-passing layers. After node updates, a readout phase typically aggregates the updated node embeddings to construct a graph-level representation, which is then passed to MLPs for property prediction. Here, we evaluated several conventional GNNs, including Graph Convolutional Networks (\textbf{GCN}) \cite{kipf2017semisupervised}, Graph Attention Networks (\textbf{GATv2}) \cite{brody2022how}, Graph Isomorphism Network (\textbf{GIN}) \cite{xu2018how}, and \textbf{AttentiveFP} \cite{Xiong2020}. We also tested more recent architectures, such as Principal Neighbourhood Aggregation (\textbf{PNA}) \cite{PNA}, Graph Transformers (\textbf{GT}) \cite{ijcai2021p214}, and \textbf{GPS} \cite{GPS}.  To examine the effect of 3D structural information, we generated monomer conformers by initialising geometries using ETKDGv3 \cite{Wang2020} and optimising them with the MMFF force field \cite{halgren1996merck}. We then applied \textbf{DimeNet++} \cite{gasteiger_dimenetpp_2020}, an SE(3)-invariant network, for polymer property prediction.  

Recently proposed GNN variants that combine multiple architectures were also evaluated. These include a GATv2 layer followed by a GraphSAGE layer (\textbf{GATv2-SAGE}) \cite{Queen2023}, GATv2 followed by LineEvo layers (\textbf{GATv2-LineEvo}) \cite{Ren2023}, and KAN-based GNNs (\textbf{KAN-GCN} and \textbf{KAN-GATv2}) \cite{Li2025}, where node embedding projections were replaced with Fourier KAN layers. Furthermore, we tested replacing MLPs in GNNs with FastKAN, resulting in \textbf{FastKAN-GATv2}, \textbf{FastKAN-GIN}, and \textbf{FastKAN-GPS}. Detailed information on how KAN is incorporated into GNNs can be found in Section~S3.1 of the SI. Finally, we explored GATv2 with different polymer representations: \textbf{GATv2-Periodic} for periodic graphs and \textbf{GATv2-VN} for graphs with virtual nodes, using RDKit descriptors as the initial embeddings for the virtual nodes.

%% file: section/strategy.tex
\section{Training Strategies}
\subsection{Multi-fidelity Learning}
\label{multi-fidelity}
In polymer property prediction, it is common for datasets to originate from different sources, such as MD simulations or experimental measurements, and exhibit varying fidelities. Here, we explored three strategies for handling multi-fidelity data, using GATv2 as the base model and polymer density prediction as an example task. We used density datasets of different fidelities (MD and experimental data) and evaluated model performance using 5-fold cross-validation on the experimental dataset as the benchmark. We used the following three approaches (detailed information on these multi-fidelity strategies can be found in Section~S3.2 of the SI):

\begin{itemize}
    \item \textbf{Finetuning (transfer learning)}: we trained a GATv2 model on the low-fidelity dataset and then finetuned it on the high-fidelity dataset. The polymer representations were either frozen or updated during finetuning, referred to as finetune (freeze) and finetune (all), respectively.
    \item \textbf{Label residual}: we trained a GATv2 model on the low-fidelity dataset, then used a new GATv2 model to predict the residual between the labels estimated by the pretrained model and the ground-truth high-fidelity labels. This approach is referred to as residual (label).
    \item \textbf{Embedding residual}: we trained a GATv2 model on the low-fidelity dataset and extracted the graph embeddings from the pretrained model. A new GATv2 was then trained with these pretrained embeddings added to its graph embeddings. This approach is referred to as residual (emb).
\end{itemize}

\subsection{$\Delta$-Learning}

$\Delta$-Learning is particularly useful for multi-task and multi-fidelity learning (e.g., the label residual and embedding residual strategies described in Section~\ref{multi-fidelity}). $\Delta$-Learning also provides a convenient way to incorporate prior knowledge into GNNs. We evaluated three $\Delta$-Learning strategies as follows (detailed information can be found in Section~S3.3 of the SI):

\begin{itemize}
    \item \textbf{Property knowledge transfer}: Many polymer properties are correlated. To leverage information from related properties, we incorporated graph embeddings from a GATv2 model pretrained on other properties.
    
    \item \textbf{Empirical equations}: Certain properties can be roughly estimated using empirical equations. To exploit this prior knowledge, we trained a GATv2 model to predict the residual between the ground-truth labels and the estimates from these equations. For polymer density, we implemented three estimators: a van der Waals-based estimator (vdW), Fedors' group contribution method (Fedors) \cite{fedors}, and a group contribution method from the IBM GitHub repository (IBM) \cite{IBM_polymer_property_prediction}. For the radius of gyration ($R_{g}$), we constructed an estimator based on the monomer and a predefined chain length (mono). Additional details are provided in Section S3.3.2 of the SI.
    
    \item \textbf{Atomic contribution}: Similar to approaches used in energy prediction, simple atomic contribution models can provide rough property estimates based on atom counts. We trained a GATv2 model to learn the residual between the ground-truth labels and these atomic-based estimates.
\end{itemize}


\subsection{Active Learning}
Using $R_{g}$ as the example, we introduce the MD setup and the active learning strategy used here. The initial $R_{g}$ dataset was compiled from two prior studies by Hayashi $et al.$ \cite{Hayashi2022} and Liu $et$ $al.$ \cite{neurips-open-polymer-prediction-2025}, both of which reported $R_{g}$ values obtained from MD simulations. The dataset from Liu $et$ $al.$\cite{neurips-open-polymer-prediction-2025} contains both labelled and unlabelled polymers. The labelled subset, along with those from Hayashi's dataset, was used to train the initial model, while the unlabelled subset served as a candidate pool for active learning. Active learning is designed to improve predictive accuracy by selectively sampling new data points that are expected to be most informative. Here, we applied a standard active learning workflow to demonstrate how our unified framework can systematically improve polymer property prediction. The workflow consists of the following steps:

\begin{enumerate}[label=\arabic*)]
    \item \textbf{Initial training}: A GATv2 model was trained on the initial labelled dataset. Scaffold-based splitting was used to create training, validation, and test sets, and model hyperparameters were optimised using the validation set. The optimised hyperparameters were fixed and reused throughout all active importing cycles. The scaffold-based test set was held out and used only for evaluation during the active learning process.
    \item \textbf{Data acquisition}: Two acquisition strategies were compared: uncertainty-based sampling and random sampling. For uncertainty-based sampling, a model was trained using 5-fold cross-validation on the combination of the training set and validation set and then used to predict $R_{g}$ values for all unlabelled monomers. Prediction uncertainty was quantified as the standard deviation of $R_{g}$ predictions across the five models. The 100 monomers with the highest uncertainty were selected for labelling. For random sampling, 100 monomers were selected uniformly at random from the unlabelled pool without using a model. All selected monomers were labelled by running new MD simulations.
    \item \textbf{Model retraining}: The newly labelled samples were added to the existing training dataset, and the model was retrained using the same 5-fold cross-validation procedure.
    \item \textbf{Evaluation}: Model performance was evaluated on the fixed hold-out test set from scaffold splitting using mean absolute error (MAE).
\end{enumerate}

Each pass through these steps constitutes one active learning round. The process was repeated for six rounds, resulting in a total of 600 newly labelled and unique monomers. MD simulations were performed with RadonPy\cite{Hayashi2022} to generate labels for selected monomers. Further details are in Section S3.4 of the SI.


\subsection{Ensemble Learning}
Ensemble learning is a common strategy for improving predictive accuracy and robustness by combining multiple models rather than relying on a single predictor.
PolyMon incorporates several ensemble learning strategies to systematically assess their impact on polymer property prediction. Specifically, we implemented \textbf{voting}, \textbf{bagging}, \textbf{gradient boosting}\cite{gradientboosting}, \textbf{snapshot}\cite{huang2017snapshot}, and \textbf{soft gradient boosting}\cite{softgradient}, following standard formulations adapted from the TorchEnsemble GitHub repository\cite{TorchEnsemble_Ensemble_PyTorch}.
 These methods are supported within the same training and evaluation pipeline, enabling direct comparison with non-ensemble models. $R_{g}$ was used as a representative case to demonstrate the utility of ensemble learning. Ensemble models were trained on the same initial scaffold-split dataset used in the active learning experiments, while evaluation was performed on the corresponding hold-out test set using MAE. To examine the effect of ensemble size, we considered four numbers of base estimators, including 5, 10, 15, and 20, for each ensemble method. For reference, a single model trained without ensembling was included as a baseline. All models, both with and without ensembles, were trained using identical hyperparameters to ensure a fair comparison. 

%% file: section/result.tex
\section{Results and discussion}
\subsection{Datasets and metrics}
We investigated different methods for predicting five polymer properties: glass transition temperature ($T_{g}$), fractional free volume (FFV), thermal conductivity (TC), density, and radius of gyration ($R_{g}$). These properties are critical for various polymer applications, including as membranes and electrolytes. As summarised in Table~\ref{dataset}, most datasets contained over one thousand entries. The $T_{g}$ and \(\rho_{\rm exp}\) datasets are experimentally measured, while the remaining datasets (FFV, Tc, $R_{g}$, and \(\rho_{\rm md}\)) are derived from MD simulations.

\begin{table}[h!]
\centering
\caption{Dataset summary}
\label{dataset}
\begin{tabular}{l|lll|l}
\toprule
Dataset& Description & Unit & \# data points & Source \\
\midrule
$T_{g}$ & Glass transition temperature& $^\circ$C & 7,367 & \cite{kunchapu2025polymetrix}\\
FFV &Fractional free volume & 1 & 7,030 & \cite{neurips-open-polymer-prediction-2025}\\
TC & Thermal conductivity& $\rm W/(m\cdot K)$ & 1,070 & \cite{Hayashi2022}\\
$\rho_{\rm md}$ & MD-calculated density& $\rm g/cm^{3}$ & 1,077 & \cite{Hayashi2022} \\
$R_{g}$ & Radius of gyration &  \AA & 1,328 & \cite{Hayashi2022, neurips-open-polymer-prediction-2025}  \\
$\rho_{\rm exp}$ & Experimentally measured density & $\rm g / cm^{3}$ & 117 & \cite{Afzal2021}\\
\bottomrule
\end{tabular}
\end{table}

We used $T_{g}$, FFV, TC, \(\rho_{\rm md}\), and $R_{g}$ to evaluate the performance of descriptor-based tabular models, graph-based GNNs, and $\Delta$-learning approaches. MAE was employed as the primary metric for comparing different methods across datasets. The datasets \(\rho_{\rm exp}\) and \(\rho_{\rm md}\) were used to assess the performance of various multi-fidelity training strategies. To reduce the influence of dataset splitting, we report the mean MAE over 5-fold cross-validation as the final performance. Additionally, we used the weighted MAE (wMAE) \cite{neurips-open-polymer-prediction-2025} to quantify overall performance across the five properties. The wMAE is calculated as:

\begin{equation}
    \text{wMAE}=\frac{1}{|\mathcal{X}|}\sum_{X\in\mathcal{X}}\sum_{i\in\mathcal{I}(X)}\omega_i\cdot |\hat y_i(X)-{y_i}(X)|,
\end{equation}
where $\mathcal{X}$ is the set of polymers being evaluated, and $\mathcal{I}(X)$ is the property sets for polymer $X$. The terms $\hat y_i(X)$ and  $y_i(X)$ are the predicted label and ground truth label for the $i$-th property of polymer $X$. The weights $w_i$ for different properties are calculated by:
\begin{equation}
    w_i=\frac{1}{(y_i^{\text{max}}-y_i^{\text{min}})}\cdot\frac{K\cdot\sqrt{1/n_i}}{\sum_{j=1}^K\sqrt{1/n_j}},
\end{equation}
where $n_i$ is the number of data points for the $i$-th property, and $K$ is the number of property sets ($K=5$ in this work). For active learning and ensemble learning, we kept a hold-out test set and used the MAE on the test set as the performance.

\subsection{The performance of tabular models}
We first evaluated various tabular models using RDKit descriptors. As shown in Table \ref{tab:tabular_model}, TabPFN emerged as the top performer, achieving the lowest wMAE of 0.0228 and consistently delivering the best results across all individual properties. Benefiting from pretraining on millions of synthetic datasets, TabPFN effectively captures both property–feature relationships and inter-feature dependencies. The KAN variants, specifically FastKAN and EfficientKAN, ranked as the second-best architectures, outperforming MLP and traditional tree-based models such as RF and XGBoost. Owing to their learnable activation functions, KANs possess stronger expressive capacity than standard MLPs.

\input{table/tabular_models}

We further evaluated the impact of different molecular descriptors using an RF baseline (Table \ref{tab:rf_descriptors}). Both Mordred (dimer) and standard Mordred descriptors achieved the highest predictive accuracy, with wMAEs of 0.0286 and 0.0294, respectively. RDKit (dimer) and standard RDKit descriptors yielded the second-best performance. Mordred and RDKit offer a diverse set of descriptors, including functional group counts, topological features, and estimated properties, which collectively help tabular models capture the structure-property relationships. Interestingly, descriptors computed on dimers often perform as well as, or better than, those computed on monomers. This improvement is to be expected and likely arises because dimers encode additional information, including the inter-monomer relationships. Notably, while the ECFP4 fingerprint is generally less effective for most thermodynamic properties, it delivered the best prediction for the radius of gyration ($\text{$R_{g}$ }= 2.289 \pm 0.124$), highlighting that the structural connectivity captured by fingerprints is particularly relevant for polymer chain dimensions. Pretrained embeddings from language models (PolyCL and PolyBERT), on the other hand, did not outperform the conventional descriptors. These embeddings are obtained through self-supervised learning on chemical structures; however, their non-end-to-end nature may limit their task-specific performance. We also evaluated other combinations of models and descriptors, with the results presented in Section~S4.1 of the SI.

\input{table/descriptor}

\subsection{Performance of GNNs}

We evaluated the performance of different GNN architectures for polymer property prediction (Table~\ref{tab:dl_models}). Among the tested models, the PNA network achieved the best overall performance with a wMAE of 0.0193. The GPS architecture also showed highly competitive results, particularly for predicting $T_{g}$ and $R_{g}$. GPS combines a global attention layer (graph transformer) to capture long-range interactions with a local message-passing layer for short-range interactions, which likely contributes to its strong performance. The GT model also performed well, highlighting the importance of incorporating long-range interactions in polymer property prediction.  

Other classic GNNs, including GCN, GATv2, AttentiveFP, and GIN, maintained high accuracy, demonstrating the effectiveness of graph-based representations in capturing complex structure-property relationships. Conversely, the KAN- and FastKAN-based GNNs did not outperform their standard counterparts, suggesting that further careful network design is required to fully leverage the potential of KAN architectures. Moreover, although 3D structural information is generally important for polymer properties, DimeNet++ performed poorly, likely because the force-field-generated monomer structures were insufficiently accurate.  

We also tested two GATv2 variants augmented with GraphSAGE and LineEvo layers. Both variants outperformed the vanilla GATv2, consistent with findings reported in \citet{Queen2023, Ren2023}. This improvement may result from neighbour sampling in GraphSAGE and the inclusion of coarse-grained information in LineEvo. Regarding graph construction strategies, periodic graphs outperformed monomer graphs, likely because they allow message passing between attachment sites. In contrast, GATv2-VN (graphs with virtual nodes) underperformed relative to the other representations, especially for $T_{g}$, showing larger errors and standard deviations. We also tested three additional GNNs on certain tasks, including two novel methods proposed in this work. These models either underperformed relative to the baseline or could not complete within the specified time; their performance is therefore reported in Section~S4.2 of the SI.

To compare the performance of GNNs with tabular models, we evaluated TabPFN and LightGBM using Mordred (dimer) descriptors and compared them with PNA and GPS. As shown in Figure~\ref{fig:dl_vs_ml}, for most properties except density, PNA and GPS outperform TabPFN and LightGBM. Interestingly, TabPFN achieves comparable or even superior performance to some GNNs, highlighting that descriptor-based tabular models remain competitive for polymer property prediction.

\input{table/gnn}

\begin{figure*}[h]
    \centering
	\includegraphics[width=\linewidth]{figure/dl_vs_ml.png}
	\caption{Performance comparison between tabular models (TabPFN and LightGBM) and GNNs (PNA and GPS) on five polymer properties.}
	\label{fig:dl_vs_ml}
\end{figure*}

\subsection{The performance of different training strategies}
\begin{figure*}[h]
    \centering
	\includegraphics[width=\linewidth]{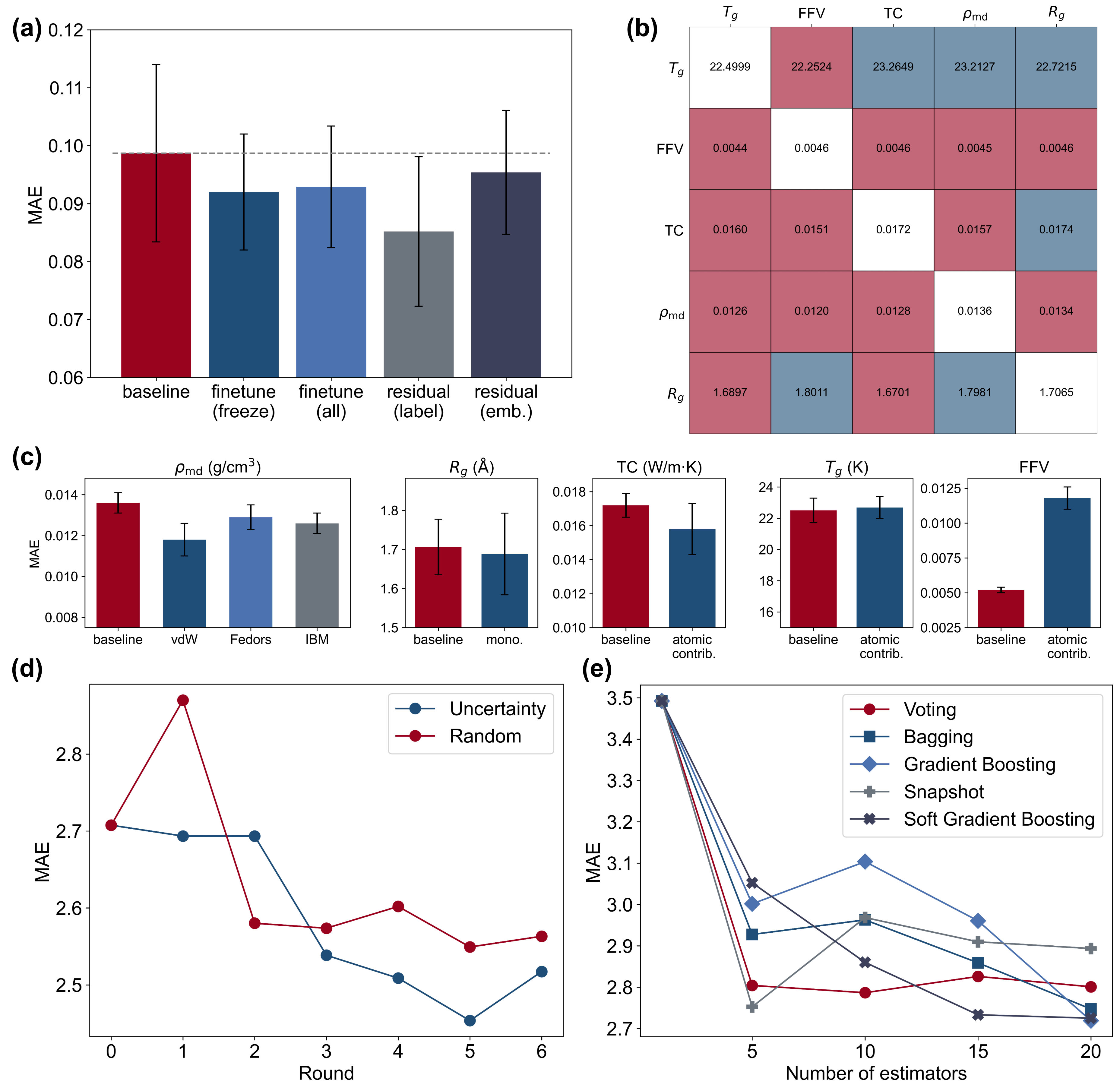}
	\caption{
        Performance of different training strategies. 
        (a) Multi-fidelity learning on the experimental density dataset (\(\rho_{\rm exp}\)); baseline: GATv2 trained directly on \(\rho_{\rm exp}\). \textbf{finetune (freeze/all)}: pretrained on \(\rho_{\rm md}\) then finetuned on \(\rho_{\rm exp}\), with frozen or fully trainable parameters. \textbf{residual (label/emb)}: residual learning on labels or graph embeddings.  
        (b) Property knowledge transfer: each row shows the target property using embeddings from the column property. Diagonal: baseline; red: improved, blue: worse.  
        (c) $\Delta$-learning using empirical equations (\(\rho_{\rm md}\), $R_{g}$) or atomic contributions TC, $T_{g}$, FFV).  
        (d) Active learning: "Random" selects points randomly; "Uncertainty" selects based on uncertainty.  
        (e) Ensemble performance with varying numbers of estimators.
}
	\label{fig:training_strategy}
\end{figure*}

As shown in Fig.~\ref{fig:training_strategy}a, we evaluated the performance of different strategies for the multi-fidelity task on the density property. All strategies achieved better performance than the baseline, which corresponds to a GATv2 trained directly on \(\rho_{\rm exp}\). For the finetuning strategy, we tested two cases: freezing the GNN layers and training only the prediction head (finetune (freeze)) and training all parameters during finetuning (finetune (all)). Both approaches achieved similar performance, with finetune (freeze) performing slightly better, likely due to reduced forgetting compared to finetune (all). For the residual strategy, we compared learning the residual at the label level (residual (label)) versus the embedding level (residual (emb)). We found that using the estimated labels from the low-fidelity model provides more improvement for the high-fidelity model than using the low-fidelity embeddings. In particular, the label residual strategy improved performance by over 10\% relative to the baseline model.

$\Delta$-learning has been widely used for training ML force fields \cite{doi:10.1021/acs.jctc.2c01034}. Here, we found that it can also improve performance in polymer property prediction (Figs.~\ref{fig:training_strategy}b and c). The performance matrix for the property knowledge transfer strategy shows that the prediction of one property can be enhanced by incorporating graph embeddings from other properties, particularly for the TC, \(\rho_{\rm md}\), and $R_{g}$ datasets. These datasets are relatively small compared to $T_{g}$ and FFV, and can benefit from the inductive bias provided by related properties. This demonstrates that property knowledge transfer can effectively mitigate data scarcity for certain polymer properties. $\Delta$-learning on labels is another effective strategy to introduce inductive bias or reduce training difficulty. As shown in Fig.~\ref{fig:training_strategy}c, all density estimation methods improved performance relative to the baseline, with the vdW-based estimator achieving the best results. Interestingly, although the vdW-based estimator does not provide more accurate estimates than the two other group contribution-based methods (Figure~S1), it helps GNNs achieve better performance. This may be because the group contribution methods can produce inconsistent estimates when certain functional groups are absent from their predefined sets, reducing stability. In contrast, the simple vdW-based estimator offers consistent and stable predictions, even if it is less accurate than the group contribution methods. Similar trends are observed for $R_{g}$, where the monomer-based estimator (mono.) improves performance compared to the baseline. Atomic contribution methods, however, sometimes degrade performance, as seen for $T_{g}$ and FFV. For the TC dataset, the atomic contribution method improves performance by approximately 8\%.

Active learning offers a promising approach to improve the performance of property prediction models with high data efficiency. Fig.~\ref{fig:training_strategy}d compares uncertainty-guided data selection (i.e., active learning) with random data selection. By identifying out-of-distribution points, uncertainty-based selection can improve model generalisability and achieve better performance with fewer labelled data compared to random selection. As the number of active learning rounds increases, both strategies show improved performance, highlighting the importance of incorporating additional labelled data. Combining active learning with MD simulations represents a particularly promising direction for polymer property prediction.

Ensemble learning is commonly used to improve predictive performance by training multiple models and aggregating their predictions. The performance of different ensemble strategies is shown in Fig.~\ref{fig:training_strategy}e. In general, most strategies show improved performance as the number of estimators increases. Among the five ensemble strategies tested, we found that the simple voting strategy achieved the best overall performance. Compared to single models, all ensemble methods provided measurable improvements, with the best ensemble model (gradient boosting with 20 estimators) improving performance by over 20\%.


%% file: table/tabular_models.tex
\begin{table}[htbp]
\centering
\caption{MAE of tabular models using RDKit descriptors (lower is better). Note that bold values indicate the best-performing models; underlined values indicate the second best.}
\label{tab:tabular_model}
\resizebox{\textwidth}{!}{
\renewcommand{\arraystretch}{1.2}
\begin{tabular}{l|ccccc|c}
\hline
Model & $\rho_{\rm md}$ & FFV & $R_{g}$ & TC & $T_{g}$ & wMAE\\
\hline
TabPFN &
$\mathbf{0.0121}_{\pm 0.0012}$ &
$\mathbf{0.0053}_{\pm 0.0003}$ &
$\mathbf{2.0015}_{\pm 0.1536}$ &
$\mathbf{0.0171}_{\pm 0.0004}$ &
$\mathbf{23.5880}_{\pm 0.1705}$ &
$\mathbf{0.0228}$ \\

FastKAN &
$\underline{0.0146}_{\pm 0.0010}$ &
$\underline{0.0061}_{\pm 0.0002}$ &
$2.2862_{\pm 0.1948}$ &
$0.0214_{\pm 0.0027}$ &
$26.2204_{\pm 1.0159}$ &
$\underline{0.0271}$ \\

EfficientKAN &
$0.0167_{\pm 0.0014}$ &
$0.0066_{\pm 0.0001}$ &
$\underline{2.2535}_{\pm 0.2179}$ &
$\underline{0.0211}_{\pm 0.0019}$ &
$28.7122_{\pm 1.1073}$ &
$0.0279$ \\

MLP &
$0.0149_{ \pm 0.0010}$ &
$0.0065_{ \pm 0.0001}$ &
$2.3959_{ \pm 0.02641}$ &
$\underline{0.0211}_{ \pm 0.0012}$ &
$29.0279_{ \pm 0.7730}$  &
$0.0279$ \\

LightGBM &
$0.0223_{\pm 0.0031}$ &
$0.0067_{\pm 0.0004}$ &
$2.4571_{\pm 0.0969}$ &
$0.0216_{\pm 0.0011}$ &
$\underline{25.3985}_{\pm 0.3770}$ &
$0.0295$ \\

CatBoost &
$0.0229_{\pm 0.0033}$ &
$\underline{0.0061}_{\pm 0.0005}$ &
$2.3854_{\pm 0.1789}$ &
$0.0228_{\pm 0.0008}$ &
$25.9516_{\pm 0.4698}$ &
$0.0300$ \\

RF &
$0.0273_{\pm 0.0043}$ &
$0.0073_{\pm 0.0004}$ &
$2.4820_{\pm 0.1760}$ &
$0.0239_{\pm 0.0012}$ &
$28.1031_{\pm 0.4177}$ &
$0.0325$ \\

XGBoost &
$0.0283_{\pm 0.0040}$ &
$0.0071_{\pm 0.0004}$ &
$3.4709_{\pm 0.1709}$ &
$0.0235_{\pm 0.0014}$ &
$33.3332_{\pm 0.4230}$ &
$0.0365$ \\

FourierKAN &
$0.0597_{\pm 0.0050}$ &
$0.0108_{\pm 0.0003}$ &
$3.1513_{\pm 0.2595}$ &
$0.0296_{\pm 0.0022}$ &
$36.9348_{\pm 1.1130}$ &
$0.0472$ \\
\hline
\end{tabular}}
\end{table}

%% file: table/descriptor.tex
\begin{table}[htbp]
\centering
\caption{MAE of RF models with different descriptors (lower is better). Note that bold values indicate the best-performing descriptors; underlined values indicate the second best.}
\label{tab:rf_descriptors}
\resizebox{\textwidth}{!}{
\renewcommand{\arraystretch}{1.2}
\begin{tabular}{l|ccccc|c}
\hline
Descriptor & $\rho_{\rm md}$ & FFV & $R_{g}$ & TC & $T_{g}$ & wMAE\\
\hline
Mordred (dimer) &
$\mathbf{0.0203}_{\pm 0.0019}$ &
$\mathbf{0.0071}_{\pm 0.0004}$ &
$\underline{2.2941}_{\pm 0.1957}$ &
$\mathbf{0.0213}_{\pm 0.0013}$ &
$\mathbf{25.9871}_{\pm 0.4465}$ &
$\mathbf{0.0286}$ \\

Mordred &
$\underline{0.0202}_{\pm 0.0025}$ &
$\mathbf{0.0071}_{\pm 0.0004}$ &
$2.3571_{\pm 0.1424}$ &
$\underline{0.0224}_{\pm 0.0014}$ &
$\underline{26.7003}_{\pm 0.4835}$ &
$\underline{0.0294}$ \\

RDKit (dimer) &
$0.0276_{\pm 0.0035}$ &
$\underline{0.0073}_{\pm 0.0004}$ &
$2.4329_{\pm 0.1873}$ &
$0.0235_{\pm 0.0009}$ &
$27.3671_{\pm 0.3136}$ &
$0.0321$ \\

RDKit &
$0.0273_{\pm 0.0043}$ &
$\underline{0.0073}_{\pm 0.0004}$ &
$2.4820_{\pm 0.1760}$ &
$0.0239_{\pm 0.0012}$ &
$28.1031_{\pm 0.4177}$ &
$0.0325$ \\

MACCS &
$0.0347_{\pm 0.0056}$ &
$0.0087_{\pm 0.0003}$ &
$2.5062_{\pm 0.1838}$ &
$0.0255_{\pm 0.0007}$ &
$28.5440_{\pm 0.7880}$ &
$0.0355$ \\

ECFP4 &
$0.0434_{\pm 0.0046}$ &
$0.0085_{\pm 0.0004}$ &
$\mathbf{2.2886}_{\pm 0.1243}$ &
$0.0242_{\pm 0.0005}$ &
$28.3704_{\pm 0.6550}$ &
$0.0360$ \\

ECFP4 (dimer) &
$0.0443_{\pm 0.0040}$ &
$0.0087_{\pm 0.0004}$ &
$2.3258_{\pm 0.1686}$ &
$0.0245_{\pm 0.0011}$ &
$28.7794_{\pm 0.8403}$ &
$0.0366$ \\

PolyCL &
$0.0456_{\pm 0.0023}$ &
$0.0091_{\pm 0.0005}$ &
$2.5627_{\pm 0.2454}$ &
$0.0263_{\pm 0.0021}$ &
$31.9435_{\pm 0.9320}$ &
$0.0392$ \\

PolyBERT &
$0.0483_{\pm 0.0033}$ &
$0.0099_{\pm 0.0005}$ &
$2.4666_{\pm 0.1639}$ &
$0.0252_{\pm 0.0015}$ &
$33.1941_{\pm 0.5011}$ &
$0.0393$ \\
\hline
\end{tabular}}
\end{table}

%% file: table/gnn.tex
\begin{table}[H]
\centering
\caption{MAE of deep learning models on polymer properties (lower is better). Note that bold values indicate the best-performing models; underlined values indicate the second best.}
\label{tab:dl_models}
\resizebox{\textwidth}{!}{
\renewcommand{\arraystretch}{1.2}
\begin{tabular}{l|ccccc|c}
\hline
Model & $\rho_{\rm md}$ & FFV & $R_{g}$ & TC & $T_{g}$ & wMAE\\
\hline
PNA &
$\underline{0.0129}_{\pm 0.0008}$ &
$\mathbf{0.0043}_{\pm 0.0002}$ &
$\underline{1.5583}_{\pm 0.1598}$ &
$\mathbf{0.0134}_{\pm 0.0017}$ &
$22.1203_{\pm 0.7736}$ &
$\mathbf{0.0193}$ \\

GPS &
$0.0143_{\pm 0.0009}$ &
$\underline{0.0044}_{\pm 0.0002}$ &
$\mathbf{1.5377}_{\pm 0.1242}$ &
$0.0151_{\pm 0.0017}$ &
$\mathbf{21.8669}_{\pm 1.0246}$ &
$\underline{0.0204}$ \\

GCN &
$0.0122_{\pm 0.0004}$ &
$0.0046_{\pm 0.0002}$ &
$1.7321_{\pm 0.1783}$ &
$0.0155_{\pm 0.0016}$ &
$\underline{21.9623}_{\pm 1.0744}$ &
$0.0208$ \\

AttentiveFP &
$0.0122_{\pm 0.0007}$ &
$0.0046_{\pm 0.0001}$ &
$1.6858_{\pm 0.1368}$ &
$0.0162_{\pm 0.0021}$ &
$22.8237_{\pm 1.1944}$ &
$0.0211$ \\

GT &
$0.0124_{\pm 0.0008}$ &
$0.0045_{\pm 0.0002}$ &
$1.8355_{\pm 0.1740}$ &
$0.0158_{\pm 0.0015}$ &
$22.0453_{\pm 0.6657}$ &
$0.0213$ \\

GIN &
$0.0139_{\pm 0.0016}$ &
$0.0047_{\pm 0.0002}$ &
$1.6918_{\pm 0.1640}$ &
$0.0174_{\pm 0.0021}$ &
$23.1889_{\pm 0.5599}$ &
$0.0222$ \\

FastKAN-GPS &
$0.0162_{\pm 0.0020}$ &
$0.0047_{\pm 0.0002}$ &
$1.7733_{\pm 0.2368}$ &
$0.0185_{\pm 0.0025}$ &
$22.0537_{\pm 0.6651}$ &
$0.0234$ \\

KAN-GCN &
$0.0173_{\pm 0.0016}$ &
$0.0075_{\pm 0.0027}$ &
$1.7718_{\pm 0.1600}$ &
$0.0163_{\pm 0.0014}$ &
$25.7657_{\pm 0.9712}$ &
$0.0237$ \\

DimetNet++ &
$0.0266_{\pm 0.0014}$ &
$0.0085_{\pm 0.0006}$ &
$1.8707_{\pm 0.1656}$ &
$0.0216_{\pm 0.0015}$ &
$30.5145_{\pm 1.1085}$ &
$0.0298$ \\

FastKAN-GIN &
$0.0328_{\pm 0.0025}$ &
$0.0060_{\pm 0.0006}$ &
$3.2891_{\pm 0.4014}$ &
$0.0290_{\pm 0.0113}$ &
$41.2818_{\pm 2.0626}$ &
$0.0409$ \\

\hline

GATv2-SAGE &
$\mathbf{0.0121}_{\pm 0.0006}$ &
$0.0048_{\pm 0.0004}$ &
$1.6249_{\pm 0.0863}$ &
$0.0156_{\pm 0.0008}$ &
$22.4938_{\pm 0.9094}$ &
$0.0206$ \\

GATv2-LineEvo &
$0.0130_{\pm 0.0006}$ &
$0.0048_{\pm 0.0001}$ &
$1.6383_{\pm 0.1731}$ &
$\underline{0.0150}_{\pm 0.0019}$ &
$23.3476_{\pm 0.4781}$ &
$0.0207$ \\

GATv2 &
$0.0136_{\pm 0.0005}$ &
$0.0046_{\pm 0.0001}$ &
$1.7065_{\pm 0.0711}$ &
$0.0172_{\pm 0.0007}$ &
$22.4999_{\pm 0.7899}$ &
$0.0220$ \\

FastKAN-GATv2 &
$0.0138_{\pm 0.0012}$ &
$0.0048_{\pm 0.0002}$ &
$1.7773_{\pm 0.1565}$ &
$0.0163_{\pm 0.0015}$ &
$24.8758_{\pm 1.0911}$ &
$0.0222$ \\

KAN-GATv2 &
$0.0195_{\pm 0.0067}$ &
$0.0066_{\pm 0.0001}$ &
$1.7889_{\pm 0.1724}$ &
$0.0161_{\pm 0.0009}$ &
$25.2952_{\pm 1.1712}$ &
$0.0239$ \\

\hline

GATv2-Periodic &
$0.0143_{\pm 0.0023}$ &
$0.0047_{\pm 0.0002}$ &
$1.6690_{\pm 0.1459}$ &
$0.0155_{\pm 0.0017}$ &
$23.3959_{\pm 0.8349}$ &
$0.0213$ \\

GATv2-VN &
$0.0162_{\pm 0.0009}$ &
$0.0049_{\pm 0.0002}$ &
$2.4094_{\pm 0.2087}$ &
$0.0213_{\pm 0.0020}$ &
$58.8836_{\pm 31.6556}$ &
$0.0330$ \\

\hline
\end{tabular}}
\end{table}

%% file: section/conclusion.tex
\section{Conclusion}

We have developed \textbf{PolyMon}, a unified framework for polymer property prediction. PolyMon supports a range of data representations, from conventional descriptors to graph structures, and integrates diverse ML models, including tabular models, GNNs, and novel variants. Unlike other property prediction frameworks, PolyMon also incorporates flexible training strategies, such as multi-fidelity learning, $\Delta$-learning, active learning, and ensemble learning. Using five representative polymer properties $T_{g}$, TC, FFV, density, and $R_{g}$, we conducted comprehensive benchmarks to evaluate the performance of different models and strategies. Overall, GNNs generally outperform tabular models, though carefully designed descriptors allow tabular models to achieve comparable results in some cases. Experiments with different training strategies further demonstrate their effectiveness in improving predictive accuracy. PolyMon provides a flexible, efficient, and user-friendly platform for conducting such experiments and can serve as a foundation for advancing ML-driven polymer design in the future (the code can be found at \href{github.com/fate1997/polymon}{https://github.com/fate1997/polymon}).

%% file: table/arguments.tex
\begin{table}[htbp]
\centering
\small
\begin{tabular}{l l p{8cm}}
\hline
\textbf{Argument} & \textbf{Type} & \textbf{Description} \\
\hline
\texttt{--raw-csv} & str & Path to the raw CSV file. The CSV file should contain the SMILES column\\
\texttt{--sources} & str (multiple) & Sources to use for training (only when the input CSV file contains the Source column.)\\
\texttt{--tag} & str & Tag to use for training. \\
\texttt{--labels} & str (multiple) & Labels to be predict. The available choice is from the CSV columns.\\
\texttt{--feature-names} & str (multiple) & Feature names to use for training. Available features can be found in Table \ref{tab:polymon_features}. \\
\texttt{--n-trials} & int & Number of trials for hyperparameter optimisation. \\
\texttt{--out-dir} & str & Output directory. \\
\texttt{--hparams-from} & str & Path to hyperparameter file (.json, .pt, .pkl). \\
\texttt{--n-fold} & int & Number of cross-validation folds. \\
\texttt{--split-mode} & str & Data split mode (random, scaffold). \\
\texttt{--seed} & int & Random seed. \\
\texttt{--remove-hydrogens} & bool & Whether to remove hydrogens from molecules. \\
\texttt{--descriptors} & str (multiple) & Descriptors for training neural network models. \\
\texttt{--model} & str & Model type.  Available models are shown in Table \ref{tab:polymon_models}. \\
\texttt{--hidden-dim} & int & Hidden dimension size. \\
\texttt{--num-layers} & int & Number of model layers. \\
\texttt{--batch-size} & int & Training batch size. \\
\texttt{--lr} & float & Learning rate. \\
\texttt{--num-epochs} & int & Number of training epochs. \\
\texttt{--early-stopping-patience} & int & Early stopping patience. \\
\texttt{--device} & str & Training device. \\
\texttt{--run-production} & bool & Force 0.95:0.05:0.0 train:val:test split. \\
\texttt{--finetune} & bool & Whether to finetune the model. \\
\texttt{--finetune-csv-path} & str & CSV file for finetuning. \\
\texttt{--pretrained-model} & str & Path to pretrained model. \\
\texttt{--n-estimator} & int & Number of estimators. \\
\texttt{--additional-features} & str (multiple) & Additional training features. \\
\texttt{--skip-train} & bool & Skip training step. \\
\texttt{--low-fidelity-model} & str & Path to low-fidelity model. \\
\texttt{--estimator-name} & str & Estimator name for base predictions. \\
\texttt{--emb-model} & str & Model path to provide base graph embeddings. \\
\texttt{--ensemble-type} & str & Ensemble method. \\
\texttt{--train-residual} & bool & Label residual. \\
\texttt{--normalizer-type} & str & Normalization type (normalizer, log\_normalizer, none). \\
\texttt{--augmentation} & bool & Use data augmentation. \\
\hline
\end{tabular}
\caption{Arguments for the \texttt{polymon train} command.}
\end{table}

%% file: table/avail_descriptor.tex
\begin{table}[htbp]
\centering
\small
\begin{tabular}{l p{10cm}}
\hline
\textbf{Descriptor Name} & \textbf{Description} \\
\hline
\texttt{xenonpy\_atom} & XenonPy atom features \\
\texttt{cgcnn} & CGCNN atom features \\
\texttt{source} & Source labels as features \\
\texttt{z} & Atomic numbers as integers \\
\texttt{edge} & Default edge features (\texttt{bond}) \\
\texttt{bond} & Chemical bond features \\
\texttt{fully\_connected\_edges} & Fully connected edge indices \\
\texttt{periodic\_bond} & Add bonds between attachment points to the chemical bonds \\
\texttt{virtual\_bond} & Add a virtual node and bonds connected to other nodes\\
\texttt{pos} & 3D coordinates \\
\texttt{relative\_position} & The relative position of the atom to the nearest attachment \\
\texttt{seq} & SMILES sequence features \\
\texttt{desc} & Default descriptor features (\texttt{rdkit2d}) \\
\texttt{rdkit2d} & RDKit 2D descriptors \\
\texttt{ecfp4} & ECFP4 fingerprints \\
\texttt{rdkit3d} & RDKit 3D descriptors \\
\texttt{mordred} & Mordred descriptors \\
\texttt{maccs} & MACCS keys \\
\texttt{oligomer\_rdkit2d} & RDKit 2D descriptors of the oligomer (default is dimer) \\
\texttt{oligomer\_mordred} & Mordred descriptors of the oligomer (default is dimer) \\
\texttt{oligomer\_ecfp4} & ECFP4 fingerprints of the oligomer  (default is dimer)\\
\texttt{xenonpy\_desc} & XenonPy composition descriptors \\
\texttt{mordred3d} & Mordred 3D descriptors \\
\texttt{fedors\_density} & Estimated density from Fedors method \\
\texttt{monomer} & Monomer without attachment points \\
\texttt{polycl} & Pretrained Polycl embeddings \\
\texttt{polybert} & Pretrained PolyBERT embeddings \\
\texttt{gaff2\_mod} & Pretrained GAFF2 descriptors \\
\hline
\end{tabular}
\caption{Available feature names in PolyMon}
\label{tab:polymon_features}
\end{table}

%% file: table/avail_model.tex
\begin{table}[htbp]
\centering
\small
\begin{tabular}{p{3cm} p{4cm} p{8cm}}
\hline
\textbf{Model Type} & \textbf{Model Name} & \textbf{Description} \\
\hline
\multirow{27}{*}{GNNs} 
& \texttt{gatv2} & \href{https://arxiv.org/abs/2105.14491}{Graph Attention Network v2} \\
& \texttt{attentivefp} & \href{https://pubs.acs.org/doi/10.1021/acs.jmedchem.9b00959}{AttentiveFP} \\
& \texttt{dimenetpp} & \href{https://arxiv.org/abs/2011.14115}{DimeNet++} \\
& \texttt{gatv2vn} & GATv2 with virtual node \\
& \texttt{gin} & \href{https://arxiv.org/abs/1810.00826}{Graph Isomorphism Network} \\
& \texttt{pna} & \href{https://arxiv.org/abs/2004.05718}{Principal Neighbourhood Aggregation} \\
& \texttt{gvp} & \href{https://arxiv.org/abs/2009.01411}{Geometric Vector Perceptron} \\
& \texttt{gatv2chainreadout} & GATv2 with chain readout \\
& \texttt{gt} & \href{https://arxiv.org/abs/2009.03509}{Graph Transformer} \\
& \texttt{kan\_gatv2} & KAN-augmented GATv2 \\
& \texttt{gps} & \href{https://arxiv.org/abs/2205.12454}{Graph GPS} \\
& \texttt{kan\_gps} & KAN-augmented GraphGPS \\
& \texttt{fastkan\_gatv2} & FastKAN-augmented GATv2 \\
& \texttt{gatv2\_lineevo} & \href{https://pubs.acs.org/doi/10.1021/acs.jcim.3c00059}{GATv2 with line evolution} \\
& \texttt{gatv2\_sage} & GATv2 with SAGE aggregation \\
& \texttt{gatv2\_source} & GATv2 for multi-fidelity/source \\
& \texttt{gatv2\_pe} & GATv2 with position encoding \\
& \texttt{gatv2\_embed\_residual} & GATv2 with embedding residuals \\
& \texttt{kan\_gin} & KAN-augmented GIN \\
& \texttt{fastkan\_gin} & FastKAN-augmented GIN \\
& \texttt{kan\_gcn} & KAN-augmented GCN \\
& \texttt{dmpnn} & \href{https://pubs.acs.org/doi/full/10.1021/acs.jcim.9b00237}{Directed Message Passing Neural Network} \\
& \texttt{kan\_dmpnn} & KAN-augmented DMPNN \\
\hline
\multirow{8}{*}{Tabular} 
& \texttt{fastkan} & \href{https://github.com/ZiyaoLi/fast-kan}{Fast KAN for descriptors} \\
& \texttt{efficientkan} & \href{https://github.com/Blealtan/efficient-kan}{Efficient KAN for descriptors} \\
& \texttt{fourierkan} & \href{https://github.com/GistNoesis/FourierKAN}{Fourier KAN for descriptors} \\
& \texttt{random\_forest} & Random Forest model for tabular data \\
& \texttt{xgb} & XGBoost model for tabular data \\
& \texttt{lgbm} & LightGBM model for tabular data \\
& \texttt{catboost} & CatBoost model for tabular data \\
& \texttt{tabpfn} & TabPFN model for tabular data \\
\hline
\end{tabular}
\caption{Available models in PolyMon.}
\label{tab:polymon_models}
\end{table}

%% file: table/hparams_tabular.tex
\begin{table}[ht]
\centering
\caption{Hyperparameter search space for XGBoost.}
\label{tab:hparams_xgb}
\begin{tabularx}{\textwidth}{l X X}
\toprule
Hyperparameter & Description & Search Space \\
\midrule
\verb|n_estimators| & Number of estimators & 1000 \\
\verb|learning_rate| & Boosting learning rate & $\log$-uniform $(10^{-3}, 10^{-1})$ \\
\verb|max_depth| & Maximum tree depth & Integer $[1, 10]$ \\
\verb|subsample| & Row subsampling ratio & Uniform $(0.05, 1.0)$ \\
\verb|colsample_bytree| & Feature subsampling ratio & Uniform $(0.05, 1.0)$ \\
\verb|min_child_weight| & Minimum child weight & Integer $[1, 20]$ \\
\bottomrule
\end{tabularx}
\end{table}

\begin{table}[ht]
\centering
\caption{Hyperparameter search space for Random Forest.}
\label{tab:hparams_rf}
\begin{tabularx}{\textwidth}{l X X}
\toprule
Hyperparameter & Description & Search Space \\
\midrule
\verb|n_estimators| & Number of trees & Integer $[100, 1000]$ \\
\verb|max_depth| & Maximum tree depth & Integer $[2, 32]$ \\
\verb|min_samples_split| & Minimum samples required to split a node & Integer $[2, 10]$ \\
\verb|min_samples_leaf| & Minimum samples required at a leaf & Integer $[1, 10]$ \\
\bottomrule
\end{tabularx}
\end{table}

\begin{table}[ht]
\centering
\caption{Hyperparameter search space for LightGBM.}
\label{tab:hparams_lgbm}
\begin{tabularx}{\textwidth}{l X X}
\toprule
Hyperparameter & Description & Search Space \\
\midrule
\verb|n_estimators| & Number of estimators & Integer $[1000, 10000]$ \\
\verb|learning_rate| & Boosting learning rate & $\{0.006,0.008,0.01,0.014,0.017,0.02\}$ \\
\verb|max_depth| & Maximum tree depth & {10, 20, 100} \\
\verb|num_leaves| & Number of leaves & Integer $[1, 1000]$ \\
\verb|reg_alpha| & L1 regularization strength & $\log$-uniform $(10^{-3}, 10)$ \\
\verb|reg_lambda| & L2 regularization strength & $\log$-uniform $(10^{-3}, 10)$ \\
\verb|subsample| & Row subsampling ratio & $\{0.4,0.5,0.6,0.7,0.8,1.0\}$ \\
\verb|colsample_bytree| & Feature subsampling ratio & $\{0.3,0.4,0.5,0.6,0.7,0.8,0.9,1.0\}$ \\
\verb|min_child_samples| & Minimum data per leaf & Integer $[1, 300]$ \\
\verb|cat_smooth| & Minimum of data per categorical group & Integer $[1, 100]$ \\
\bottomrule
\end{tabularx}
\end{table}

\begin{table}[ht]
\centering
\caption{Hyperparameter search space for CatBoost.}
\label{tab:hparams_catboost}
\begin{tabularx}{\textwidth}{l X X}
\toprule
Hyperparameter & Description & Search Space \\
\midrule
\verb|learning_rate| & Learning rate & Discrete grid: $[0.001, 0.020]$ with step $0.001$ \\
\verb|depth| & Tree depth & Integer $[9, 15]$ \\
\verb|l2_leaf_reg| & L2 regularization strength & Discrete grid: $[1.0, 5.5]$ with step $0.5$ \\
\verb|min_child_samples| & Minimum samples per leaf & $\{1,4,8,16,32\}$ \\
\verb|grow_policy| & Tree growing policy & Depthwise \\
\verb|iterations| & Number of iterations for training & 10000 \\
\bottomrule
\end{tabularx}
\end{table}

\begin{table}[ht]
\centering
\caption{Hyperparameter search space for TabPFN.}
\label{tab:hparams_tabpfn}
\begin{tabularx}{\textwidth}{l X X}
\toprule
Hyperparameter & Description & Search Space \\
\midrule
\verb|n_estimators| & Number of ensemble estimators & Integer $[8, 64]$ \\
\verb|softmax_temperature| & Softmax temperature & $\{0.75, 0.8, 0.85, 0.9, 0.95, 1.0\}$ \\
\verb|average_before_softmax| & Logit averaging strategy & \{True, False\} \\
\bottomrule
\end{tabularx}
\end{table}

\begin{table}[ht]
\centering
\caption{Hyperparameter search space for FastKAN.}
\label{tab:hparams_fastkan}
\begin{tabularx}{\textwidth}{l X X}
\toprule
Hyperparameter & Description & Search Space \\
\midrule
\verb|hidden_dim| &
Hidden feature dimension &
Discrete grid: $[32, 512]$ with step $32$ \\

\verb|num_layers| &
Number of hidden layers in the FastKAN network &
Integer:$[2, 5]$ \\

\verb|grid_min| &
Minimum value of the spline grid range &
Integer"$[-5.0, -2.0]$ \\

\verb|grid_max| &
Maximum value of the spline grid range &
Integer: $[2.0, 5.0]$ \\

\verb|num_grids| &
Number of spline interpolation grids &
Discrete grid: $[6, 10]$ with step $2$ \\
\bottomrule
\end{tabularx}
\end{table}

\begin{table}[ht]
\centering
\caption{Hyperparameter search space for FourierKAN.}
\label{tab:hparams_fourierkan}
\begin{tabularx}{\textwidth}{l X X}
\toprule
Hyperparameter & Description & Search Space \\
\midrule
\verb|hidden_dim| &
Hidden feature dimension &
Discrete grid: $[32, 512]$ with step $32$ \\

\verb|num_layers| &
Number of hidden layers in the FastKAN network &
Integer:$[2, 5]$ \\

\bottomrule
\end{tabularx}
\end{table}

\begin{table}[ht]
\centering
\caption{Hyperparameter search space for EfficientKAN.}
\label{tab:hparams_efficientkan}
\begin{tabularx}{\textwidth}{l X X}
\toprule
Hyperparameter & Description & Search Space \\
\midrule
\verb|hidden_dim| &
Hidden feature dimension &
Discrete grid: $[32, 512]$ with step $32$ \\

\verb|num_layers| &
Number of hidden layers in the FastKAN network &
Integer:$[2, 5]$ \\
\bottomrule
\end{tabularx}
\end{table}

%% file: table/hparams_gnn.tex
\begin{table}[ht]
\centering
\caption{Hyperparameter search space for GATv2.}
\label{tab:hparams_gatv2}
\begin{tabularx}{\textwidth}{l X X}
\toprule
Hyperparameter & Description & Search Space \\
\midrule
\verb|hidden_dim| & Hidden feature dimension & $\{16,32,48,64\}$ \\
\verb|num_layers| & Number of graph attention layers & Integer $[2, 8]$ \\
\verb|num_heads| & Number of attention heads & $\{4, 8\}$ \\
\verb|pred_hidden_dim| & Hidden dimension of the prediction head & Distance grid:$[16,256]$ with step $16$ \\
\verb|pred_dropout| & Dropout rate in the prediction head &  $\{0.1,0.2,0.3,0.4,0.5\}$ \\
\verb|pred_layers| & Number of layers in the prediction head & $\{1,2,3\}$ \\
\bottomrule
\end{tabularx}
\end{table}

\begin{table}[ht]
\centering
\caption{Hyperparameter search space for the Graph Isomorphism Network (GIN).}
\label{tab:hparams_gin}
\begin{tabularx}{\textwidth}{l X X}
\toprule
Hyperparameter & Description & Search Space \\
\midrule
\verb|hidden_dim| &
Hidden feature dimension &
Discrete grid: $[16, 256]$ with step $16$ \\

\verb|num_layers| &
Number of GIN layers &
$\{2,3,4\}$ \\

\verb|pred_hidden_dim| &
Hidden dimension of the prediction (readout) head &
Discrete grid: $[16, 256]$ with step $16$ \\

\verb|pred_dropout| &
Dropout rate applied to the prediction head & $\{0.0,0.1,0.2,0.3,0.4,0.5\}$ \\

\verb|pred_layers| &
Number of layers in the prediction head & $\{1,2,3\}$ \\

\verb|n_mlp_layers| &
Number of MLP layers per GIN block & $\{1,2,3\}$ \\

\verb|dropout| &
Dropout rate applied to node features & $\{0.0,0.1,0.2,0.3,0.4,0.5\}$ \\

\bottomrule
\end{tabularx}
\end{table}

\begin{table}[ht]
\centering
\caption{Hyperparameter search space for the Graph Convolutional Network (GCN).}
\label{tab:hparams_gcn}
\begin{tabularx}{\textwidth}{l X X}
\toprule
Hyperparameter & Description & Search Space \\
\midrule
\verb|in_channels| & Dimension of input node feature vectors & Distance grid: $[16, 256]$ with step $16$ \\
\verb|hidden_dim| &
Hidden feature dimension &
Discrete grid: $[16, 256]$ with step $16$ \\

\verb|num_layers| &
Number of GCN layers &
$\{2,3,4\}$ \\

\verb|dropout| &
Dropout rate applied to node features & $\{0.0,0.1,0.2,0.3,0.4,0.5\}$ \\

\bottomrule
\end{tabularx}
\end{table}

\begin{table}[ht]
\centering
\caption{Hyperparameter search space for PNA.}
\label{tab:hparams_pna}
\begin{tabularx}{\textwidth}{l X X}
\toprule
Hyperparameter & Description & Search Space \\
\midrule
\verb|hidden_dim| &
Hidden feature dimension &
Discrete grid: $[16, 256]$ with step $16$ \\
\verb|towers| & Number of aggregation towers & $\{1,2,4,8\}$ \\
\verb|num_layers| & Number of PNA layers & Integer $[2, 4]$ \\
\verb|pred_hidden_dim| &
Hidden dimension of the prediction (readout) head &
Discrete grid: $[16, 256]$ with step $16$ \\
\verb|pred_dropout| &
Dropout rate applied to the prediction head &
$[0.0,0.1,0.2,0.3,0.4,0.5]$ \\
\verb|pred_layers| &
Number of layers in the prediction head & $[1,2,3]$ \\
\bottomrule
\end{tabularx}
\end{table}

\begin{table}[ht]
\centering
\caption{Hyperparameter search space for DimeNet++.}
\label{tab:hparams_dimenetpp}
\begin{tabularx}{\textwidth}{l X X}
\toprule
Hyperparameter & Description & Search Space \\
\midrule
\verb|hidden_dim| & Hidden feature dimension & Discrete grid: $[16, 256]$ with step $16$ \\
\verb|num_layers| & Number of interaction layers & Integer $[2, 4]$ \\
\verb|num_spherical| & Number of spherical basis functions & Integer $[5, 10]$ \\
\verb|num_radial| & Number of radial basis functions & $\{4,6,8,10,12\}$ \\
\verb|cutoff| & Distance cutoff for message passing & $\{2.0,2.5,3.0,3.5,4.0,4.5,5.0\}$ \\
\verb|int_emb_size| &
Dimension of interaction embedding layers &
Discrete grid: $[8, 64]$ with step $8$ \\
\verb|basis_emb_size| &
Dimension of basis function embeddings & $[4,8,12,16]$ \\
\verb|out_emb_channels| &
Dimension of output embedding channels &
Discrete grid: $[16, 256]$ with step $16$ \\
\verb|max_num_neighbors| &
Maximum number of neighbours considered per node &
Discrete grid: $[16, 64]$ with step $1$ \\
\bottomrule
\end{tabularx}
\end{table}

\begin{table}[ht]
\centering
\caption{Hyperparameter search space for AttentiveFP.}
\label{tab:hparams_attentivefp}
\begin{tabularx}{\textwidth}{l X X}
\toprule
Hyperparameter & Description & Search Space \\
\midrule
\verb|in_channels| &
Dimension of input node feature vectors &
Discrete grid: $[16, 256]$ with step $16$ \\

\verb|hidden_dim| &
Hidden feature dimension &
Discrete grid: $[16, 256]$ with step $16$ \\

\verb|num_layers| &
Number of message-passing layers &
$\{1,2,3\}$ \\

\verb|num_timesteps| &
Number of iterative attention refinement steps &
$\{1,2,3\}$ \\

\bottomrule
\end{tabularx}
\end{table}

\begin{table}[ht]
\centering
\caption{Hyperparameter search space for Graph Transformers(GT).}
\label{tab:hparams_gt}
\begin{tabularx}{\textwidth}{l X X}
\toprule
Hyperparameter & Description & Search Space \\
\midrule
\verb|hidden_dim| & Hidden feature dimension & Distance grid:$[16,512]$ with step $16$ \\
\verb|num_layers| & Number of graph attention layers & Integer $[2,6]$ \\
\verb|num_heads| & Number of attention heads & $\{4, 8\}$ \\
\verb|pred_hidden_dim| & Hidden dimension of the prediction head & Distance grid:$[128,1024]$ with step $128$ \\
\verb|pred_dropout| & Dropout rate in the prediction head &  $\{0.0,0.1,0.2,0.3,0.4,0.5\}$ \\
\verb|pred_layers| & Number of layers in the prediction head & $\{1,2,3\}$ \\
\bottomrule
\end{tabularx}
\end{table}

\begin{table}[ht]
\centering
\caption{Hyperparameter search space for the Graph Positioning System (GPS) model.}
\label{tab:hparams_gps}
\begin{tabularx}{\textwidth}{l X X}
\toprule
Hyperparameter & Description & Search Space \\
\midrule
\verb|hidden_dim| &
Hidden feature dimension &
Discrete grid: $[16, 256]$ with step $16$ \\

\verb|num_layers| &
Number of stacked GPS layers &
Discrete grid: $[2, 10]$ with step $1$ \\

\verb|pe_dim| &
Dimension of positional encodings &
$\{4,5,6,7,8\}$ \\

\verb|heads| &
Number of attention heads in the transformer component & $\{4,8\}$ \\

\verb|attn_type| &
Type of attention mechanism used in GPS &
 \{\texttt{performer}, \texttt{multihead}\} \\

\verb|attention_dropout| &
Dropout probability used inside the attention module &
$\{0.0,0.1,0.2,0.3,0.4,0.5\}$ \\

\bottomrule
\end{tabularx}
\end{table}

%% file: table/tabular_all.tex
\begin{table}[t]
\centering
\scriptsize
\setlength{\tabcolsep}{5pt}
\renewcommand{\arraystretch}{1.15}
\begin{tabular}{llccccc}
\toprule
Model & Descriptor 
& $\rho_{\rm {md}}$ 
& FFV 
& $R_{g}$
& TC 
& $T_{g}$ \\
\midrule

\multirow{10}{*}{CatBoost}
& ECFP4
& $0.0358_{\pm 0.0033}$ & $0.0072_{\pm 0.0004}$ & $2.134_{\pm 0.191}$ & $0.0214_{\pm 0.0012}$ & $23.86_{\pm 0.30}$ \\

& MACCS
& $0.0322_{\pm 0.0047}$ & $0.0082_{\pm 0.0003}$ & $2.502_{\pm 0.184}$ & $0.0243_{\pm 0.0011}$ & $28.09_{\pm 0.70}$ \\

& Mordred
& $0.0231_{\pm 0.0037}$ & $0.0059_{\pm 0.0005}$ & $2.293_{\pm 0.198}$ & $0.0199_{\pm 0.0007}$ & $23.85_{\pm 0.47}$ \\

& Oligomer-ECFP4
& $0.0372_{\pm 0.0039}$ & $0.0075_{\pm 0.0004}$ & $2.234_{\pm 0.253}$ & $0.0209_{\pm 0.0011}$ & $24.22_{\pm 0.44}$ \\

& Oligomer-Mordred
& $0.0212_{\pm 0.0037}$ & $0.0056_{\pm 0.0005}$ & $2.334_{\pm 0.232}$ & $0.0194_{\pm 0.0008}$ & $23.31_{\pm 0.46}$ \\

& Oligomer-RDKit2D
& $0.0226_{\pm 0.0030}$ & $0.0061_{\pm 0.0005}$ & $2.355_{\pm 0.221}$ & $0.0213_{\pm 0.0009}$ & $25.19_{\pm 0.42}$ \\

& PolyBERT
& $0.0431_{\pm 0.0026}$ & $0.0078_{\pm 0.0005}$ & $2.311_{\pm 0.147}$ & $0.0295_{\pm 0.0017}$ & $31.70_{\pm 0.44}$ \\

& PolyCL
& $0.0350_{\pm 0.0031}$ & $0.0073_{\pm 0.0005}$ & $2.329_{\pm 0.206}$ & $0.0226_{\pm 0.0025}$ & $27.59_{\pm 0.82}$ \\

& RDKit2D
& $0.0229_{\pm 0.0033}$ & $0.0061_{\pm 0.0005}$ & $2.385_{\pm 0.179}$ & $0.0228_{\pm 0.0008}$ & $25.95_{\pm 0.47}$ \\

& RDKit3D
& $0.1032_{\pm 0.0064}$ & $0.0197_{\pm 0.0005}$ & $3.444_{\pm 0.221}$ & $0.0354_{\pm 0.0013}$ & $68.08_{\pm 0.86}$ \\

\midrule
\multirow{10}{*}{LightGBM}
& ECFP4
& $0.0349_{\pm 0.0041}$ & $0.0073_{\pm 0.0004}$ & $2.311_{\pm 0.303}$ & $0.0222_{\pm 0.0009}$ & $23.93_{\pm 0.31}$ \\

& MACCS
& $0.0387_{\pm 0.0056}$ & $0.0083_{\pm 0.0003}$ & $2.545_{\pm 0.166}$ & $0.0263_{\pm 0.0012}$ & $30.31_{\pm 0.55}$ \\

& Mordred
& $0.0189_{\pm 0.0023}$ & $0.0057_{\pm 0.0004}$ & $2.287_{\pm 0.148}$ & $0.0209_{\pm 0.0005}$ & $23.76_{\pm 0.58}$ \\

& Oligomer-ECFP4
& $0.0410_{\pm 0.0028}$ & $0.0083_{\pm 0.0004}$ & $2.348_{\pm 0.270}$ & $0.0214_{\pm 0.0010}$ & $23.79_{\pm 0.53}$ \\

& Oligomer-Mordred
& $0.0165_{\pm 0.0018}$ & $0.0057_{\pm 0.0004}$ & $2.140_{\pm 0.212}$ & $0.0188_{\pm 0.0005}$ & $23.32_{\pm 0.45}$ \\

& Oligomer-RDKit2D
& $0.0215_{\pm 0.0031}$ & $0.0065_{\pm 0.0003}$ & $2.453_{\pm 0.201}$ & $0.0215_{\pm 0.0013}$ & $24.72_{\pm 0.33}$ \\

& PolyBERT
& $0.0394_{\pm 0.0030}$ & $0.0078_{\pm 0.0004}$ & $2.350_{\pm 0.187}$ & $0.0223_{\pm 0.0009}$ & $26.81_{\pm 0.12}$ \\

& PolyCL
& $0.0342_{\pm 0.0022}$ & $0.0072_{\pm 0.0004}$ & $2.411_{\pm 0.209}$ & $0.0220_{\pm 0.0017}$ & $26.88_{\pm 0.72}$ \\

& RDKit2D
& $0.0223_{\pm 0.0031}$ & $0.0067_{\pm 0.0004}$ & $2.457_{\pm 0.097}$ & $0.0216_{\pm 0.0011}$ & $25.40_{\pm 0.38}$ \\

& RDKit3D
& $0.0969_{\pm 0.0067}$ & $0.0198_{\pm 0.0005}$ & $3.561_{\pm 0.152}$ & $0.0361_{\pm 0.0015}$ & $68.23_{\pm 0.78}$ \\

\midrule
\multirow{10}{*}{Random Forest}
& ECFP4
& $0.0434_{\pm 0.0046}$ & $0.0085_{\pm 0.0004}$ & $2.289_{\pm 0.124}$ & $0.0242_{\pm 0.0005}$ & $28.37_{\pm 0.66}$ \\

& MACCS
& $0.0347_{\pm 0.0056}$ & $0.0087_{\pm 0.0003}$ & $2.506_{\pm 0.184}$ & $0.0255_{\pm 0.0007}$ & $28.54_{\pm 0.79}$ \\

& Mordred
& $0.0202_{\pm 0.0025}$ & $0.0071_{\pm 0.0004}$ & $2.357_{\pm 0.142}$ & $0.0224_{\pm 0.0014}$ & $26.70_{\pm 0.48}$ \\

& Oligomer-ECFP4
& $0.0443_{\pm 0.0040}$ & $0.0087_{\pm 0.0004}$ & $2.326_{\pm 0.169}$ & $0.0245_{\pm 0.0011}$ & $28.78_{\pm 0.84}$ \\

& Oligomer-Mordred
& $0.0203_{\pm 0.0019}$ & $0.0071_{\pm 0.0004}$ & $2.294_{\pm 0.196}$ & $0.0213_{\pm 0.0013}$ & $25.99_{\pm 0.45}$ \\

& Oligomer-RDKit2D
& $0.0276_{\pm 0.0035}$ & $0.0073_{\pm 0.0004}$ & $2.433_{\pm 0.187}$ & $0.0235_{\pm 0.0009}$ & $27.37_{\pm 0.31}$ \\

& PolyBERT
& $0.0483_{\pm 0.0033}$ & $0.0099_{\pm 0.0005}$ & $2.467_{\pm 0.164}$ & $0.0252_{\pm 0.0015}$ & $33.19_{\pm 0.50}$ \\

& PolyCL
& $0.0456_{\pm 0.0023}$ & $0.0091_{\pm 0.0005}$ & $2.563_{\pm 0.245}$ & $0.0263_{\pm 0.0021}$ & $31.94_{\pm 0.93}$ \\

& RDKit2D
& $0.0273_{\pm 0.0043}$ & $0.0073_{\pm 0.0004}$ & $2.482_{\pm 0.176}$ & $0.0239_{\pm 0.0012}$ & $28.10_{\pm 0.42}$ \\

& RDKit3D
& $0.1053_{\pm 0.0049}$ & $0.0195_{\pm 0.0005}$ & $3.398_{\pm 0.178}$ & $0.0358_{\pm 0.0019}$ & $69.06_{\pm 0.74}$ \\

\midrule
\multirow{10}{*}{XGBoost}
& ECFP4 
& $0.0479_{\pm 0.0041}$
& $0.0098_{\pm 0.0003}$
& $2.553_{\pm 0.140}$
& $0.0282_{\pm 0.0018}$
& $29.94_{\pm 0.63}$ \\
& MACCS 
& $0.0367_{\pm 0.0054}$
& $0.0093_{\pm 0.0003}$
& $3.460_{\pm 0.170}$
& $0.0258_{\pm 0.0012}$
& $34.89_{\pm 0.62}$ \\
& Mordred 
& $0.0209_{\pm 0.0030}$
& $0.0075_{\pm 0.0004}$
& $2.859_{\pm 0.216}$
& $0.0227_{\pm 0.0010}$
& $26.74_{\pm 0.45}$ \\
& Oligomer-ECFP4 
& $0.0441_{\pm 0.0039}$
& $0.0087_{\pm 0.0004}$
& $2.639_{\pm 0.174}$
& $0.0237_{\pm 0.0014}$
& $40.64_{\pm 0.76}$ \\
& Oligomer-Mordred 
& $0.0593_{\pm 0.0043}$
& $0.0072_{\pm 0.0004}$
& $3.598_{\pm 0.178}$
& $0.0289_{\pm 0.0019}$
& $34.80_{\pm 0.64}$ \\
& Oligomer-RDKit2D 
& $0.0417_{\pm 0.0034}$
& $0.0073_{\pm 0.0004}$
& $2.537_{\pm 0.216}$
& $0.0239_{\pm 0.0010}$
& $32.63_{\pm 0.41}$ \\
& PolyBERT 
& $0.0718_{\pm 0.0041}$
& $0.0101_{\pm 0.0005}$
& $2.542_{\pm 0.155}$
& $0.0274_{\pm 0.0010}$
& $36.76_{\pm 0.59}$ \\
& PolyCL 
& $0.0519_{\pm 0.0022}$
& $0.0096_{\pm 0.0004}$
& $2.723_{\pm 0.247}$
& $0.0303_{\pm 0.0024}$
& $36.03_{\pm 0.93}$ \\
& RDKit2D 
& $0.0283_{\pm 0.0040}$
& $0.0071_{\pm 0.0004}$
& $3.471_{\pm 0.171}$
& $0.0235_{\pm 0.0014}$
& $33.33_{\pm 0.42}$ \\
& RDKit3D 
& $0.1175_{\pm 0.0059}$
& $0.0202_{\pm 0.0006}$
& $4.354_{\pm 0.146}$
& $0.0406_{\pm 0.0017}$
& $73.38_{\pm 0.54}$ \\

\midrule
\multirow{8}{*}{TabPFN}
& MACCS 
& $0.0305_{\pm 0.0052}$
& $0.0081_{\pm 0.0003}$
& $2.278_{\pm 0.265}$
& $0.0232_{\pm 0.0007}$
& $28.46_{\pm 0.49}$ \\
& Mordred 
& $0.0118_{\pm 0.0020}$
& $0.0051_{\pm 0.0004}$
& $1.877_{\pm 0.184}$
& $0.0156_{\pm 0.0004}$
& $23.09_{\pm 0.33}$ \\
& Oligomer-Mordred 
& $0.0109_{\pm 0.0018}$
& $0.0050_{\pm 0.0004}$
& $1.747_{\pm 0.220}$
& $0.0148_{\pm 0.0007}$
& $22.75_{\pm 0.36}$ \\
& Oligomer-RDKit2D 
& $0.0116_{\pm 0.0013}$
& $0.0052_{\pm 0.0004}$
& $1.921_{\pm 0.169}$
& $0.0160_{\pm 0.0003}$
& $23.10_{\pm 0.28}$ \\
& PolyBERT 
& $0.0246_{\pm 0.0018}$
& $0.0070_{\pm 0.0003}$
& $1.909_{\pm 0.149}$
& $0.0181_{\pm 0.0006}$
& $26.29_{\pm 0.38}$ \\
& PolyCL 
& $0.0213_{\pm 0.0021}$
& $0.0065_{\pm 0.0004}$
& $2.046_{\pm 0.129}$
& $0.0182_{\pm 0.0016}$
& $26.19_{\pm 0.88}$ \\
& RDKit2D 
& $0.0121_{\pm 0.0012}$
& $0.0053_{\pm 0.0003}$
& $2.002_{\pm 0.154}$
& $0.0171_{\pm 0.0004}$
& $23.59_{\pm 0.17}$ \\
& RDKit3D 
& $0.0646_{\pm 0.0043}$
& $0.0179_{\pm 0.0006}$
& $3.112_{\pm 0.178}$
& $0.0287_{\pm 0.0011}$
& $61.27_{\pm 0.83}$ \\

\midrule
\multirow{1}{*}{FastKAN}
& RDKit2D 
& $0.0146_{\pm 0.0010}$
& $0.0061_{\pm 0.0002}$
& $2.286_{\pm 0.195}$
& $0.0214_{\pm 0.0027}$
& $26.22_{\pm 1.02}$ \\

\midrule
\multirow{1}{*}{FourierKAN}
& RDKit2D 
& $0.0597_{\pm 0.0050}$
& $0.0108_{\pm 0.0003}$
& $3.151_{\pm 0.260}$
& $0.0296_{\pm 0.0022}$
& $36.93_{\pm 1.11}$ \\

\midrule
\multirow{1}{*}{EfficientKAN}
& RDKit2D 
& $0.0167_{\pm 0.0014}$
& $0.0066_{\pm 0.0001}$
& $2.254_{\pm 0.218}$
& $0.0211_{\pm 0.0019}$
& $28.71_{\pm 1.11}$ \\

\midrule
\multirow{1}{*}{MLP}
& RDKit2D 
& $0.0149_{ \pm 0.001}$ 
& $0.0065_{ \pm 0.0001}$ 
& $2.396_{ \pm 0.0.264}$ 
& $0.021_{ \pm 0.001}$ 
& $29.03_{ \pm 0.77}$\\

\bottomrule
\end{tabular}
\caption{Prediction performance (mean$_{\pm\,\mathrm{std}}$) of different models and molecular descriptors.}
\label{tab:all_models_performance}
\end{table}